\documentclass[3p,times]{elsarticle}

\usepackage{ecrc}

\volume{00}

\firstpage{1}

\journalname{Communications in Nonlinear Science and Numerical Simulation}

\runauth{}

\jid{cnsns}

\usepackage{amssymb}
\usepackage{lineno,hyperref}
\usepackage{listing}
\usepackage{subfig}
\usepackage{todonotes}

\usepackage{amsmath,amssymb,amsfonts,amsbsy}
\usepackage{mathptmx}
\usepackage{setspace}
\usepackage{bm}

\newcommand{\bs}{\boldsymbol}

\newcommand{\intd}{\mathrm{d}\,}

\newcommand{\sens}[2]{\frac{\partial{#1}}{\partial{#2}}}
\newcommand{\drho}[1]{\frac{\partial{#1}}{\partial \rho_e}}
\newcommand{\total}[2]{\frac{\text{d}\,{#1}}{\text{d}\,{#2}}}
\newcommand{\dt}[1]{\frac{\partial{#1}}{\partial t}}

\newcommand{\bh}{{\bs h}}
\newcommand{\bbh}{{\bar{\bs h}}}

\newcommand{\bw}{{\bs w}}

\newcommand{\bK}{{\bs K}}

\newcommand{\bM}{{\bs M}}

\newcommand{\bzero}{{\bs 0}}

\newcommand{\brho}{{\bs \rho}}

\newcommand{\blmbd}{{\bs \lambda}}

\newcommand{\tK}{\widetilde{\bs{K}}}

\newcommand{\tM}{\widetilde{\bs{M}}}

\newcommand{\tabref}[1]{Tab.~\ref{#1}}
\newcommand{\figref}[1]{Fig.~\ref{#1}}
\newcommand{\secref}[1]{Sec.~\ref{#1}}

\newcommand{\eqnref}[1]{(\ref{#1})}

\begin{document}
\setlength{\parindent}{0mm}
\begin{frontmatter}



\dochead{}

\title{Topology optimization of unsaturated flows in multi-material porous media: application to a simple diaper model}


\author[opt]{Fabian Wein\corref{mycorrespondingauthor}}
\cortext[mycorrespondingauthor]{Corresponding author}
\ead{fabian.wein@fau.de}

\author[lstm]{Nan Chen}
\ead{nan.chen@fau.de}

\author[lstm]{Naveed Iqbal}
\ead{naveed.iqbal@fau.de}

\author[opt]{Michael Stingl}
\ead{michael.stingl@fau.de}

\author[lstm,zarm]{Marc Avila}
\ead{marc.avila@zarm.uni-bremen.de}




\address[opt]{Mathematical Optimization, Friedrich-Alexander-Universit\"at Erlangen-N\"urnberg, Cauerstrasse 11, 91058 Erlangen, Germany}
\address[lstm]{Institute of Fluid Mechanics, Friedrich-Alexander-Universit\"at Erlangen-N\"urnberg, Cauerstr. 4, 91058 Erlangen, Germany}
\address[zarm]{Center of Applied Space Technology and Microgravity, Universit\"at Bremen, Am Fallturm 2, 28359 Bremen, Germany}

\begin{abstract}

We present a mathematical approach to optimize the material distribution for fluid transport in unsaturated porous media. Our benchmark problem is a simplified diaper model as an exemplary liquid absorber. Our model has up to three materials with vastly different properties, which in the reference configuration are arranged in parallel layers. Here, swelling is neglected and the geometry of a swollen diaper is assumed and treated as a porous medium of high porosity. The imbibition process is then simulated by solving an unsaturated flow problem based on Richards' equation. Our aim is to maximize the amount of absorbed liquid by  redistributing the materials. To this end, a density based multi-material topology optimization (based on the well known SIMP model) is performed. A sensitivity analysis for the nonlinear transient problem is provided, which enables the application of first order optimization solvers. We perform two- and three-material optimization and discuss several variants of the problem setting. We present designs with up to 45\% more absorbed liquid compared to the reference configuration.

\end{abstract}

\begin{keyword}
\texttt Permeability \sep optimization \sep porous media \sep imbibition  
\end{keyword}

\end{frontmatter}


\section{Introduction}

Superabsorbent polymers (SAPs) are granular materials that revolutionized the market of disposable hygiene products such as diapers since their commercialization in the late seventies. The major feature of SAPs is that they can absorb and retain over thirty times their own weight of liquid while swelling. Because of this property, SAPs are widely used in sanitary products as well as in the agricultural industry. One particularly important product are modern diapers, which are composed of high-tech multi-material SAPs designed to maximize liquid absorption, while remaining dry at the contact surface with the skin. A diaper is typically composed of three parallel layers of materials with distinct properties \cite{diersch2010modeling}. The first layer is designed to evacuate the fluid of the surface as fast as possible, whereas the second one distributes the fluid in the longitudinal direction and the third is the actual SAP that stores it. 

Building upon theoretical principles derived for deformable porous media in soil mechanics \cite{lewis1998finite}, Diersch \emph{et al}.\ \cite{diersch2010modeling} derived a set of equations describing the flow of liquid and the swelling in  absorbent hygiene products. The first equation is a generalized Richards' equation extended to absorbing and swelling materials. Richards' equation \cite{Richards1931} describes the motion of liquid in an unsaturated media and finds widespread use in soil science to estimate water infiltration depth. It reduces a complex two-phase fluid problem to a partial differential equation for a single variable (the capillary pressure) and is the analogous of Darcy's equation, but for unsaturated media.  Note that in Darcy's equation pressure is the only variable, whereas in Richards' equation there is also the moisture content (saturation). For a given material there is a pressure-moisture curve, which must be measured experimentally and is typically fitted with highly nonlinear empirical laws \cite{vanGenuchten1980}. However, a grain-scale modelling technique was recently developed to predict capillary pressure-moisture curves for SAP as a function of their granular structure \cite{sweijen2016effects}. Similarly, the permeability of the material depends also nonlinearly on the saturation. The second and third equation derived by Diersch \emph{et al}.\ \cite{diersch2010modeling} are the conservation of mass in the solid phase and a relationship for the solid strain, respectively, supplemented additionally by several complex constitutive equations. While their equations model the physics of fluid absorption in SAP faithfully, simulating them accurately is a challenge. In  particular, the large swelling deformations requires moving meshes with automated refinement and several adaptive techniques must be combined to obtain robust and accurate results \cite{Diersch2011}.

An interesting question that can be numerically addressed is whether the performance of a diaper can be optimized by changing the spatial distribution of the three materials. In fact, it is common for density based topology optimization to introduce a porous model as differentiable design variable (pseudo density). Typically problems are solved where the final design consist of a topology of fully permeable and impermeable material. For example, Guest and Pr{\'e}vost \cite{Guest2006} coupled a fluid problem and topology optimization with a Darcy-Stokes approach and found out the optimal structure of porous media based on minimizing dissipated power; Shou and Fan \cite{Shou2015} investigated the fast and controlled capillary flows in homogeneous porous structure and optimized the porous structure for imbibition of capillary flow with viscous resistance;  H\"ubner et al \cite{Hubner2018} employed the Biot model for shape optimization of microstructure saturated viscous fluids and reported a two-scale structure of porous material based on effective poroelastic coefficients. However, to the best of our knowledge, multi-layer optimization of transient unsaturated flows in porous media has not been done before and is  the main focus of our work. For this purpose, we consider a highly simplified model of a diaper. The main simplification is that the swelling of the SAPs, as liquid is absorbed, is not accounted for. Instead we assume that the diaper is in its swollen state  before the absorption of liquid starts and the materials are treated as porous media. The displacement of air as the liquid fills the diaper is modeled with Richards' equation. This simplification is necessary because the complexity of the equations with swelling, and more importantly the numerical techniques necessary to solve them accurately, would render the optimization problem intractable.

In this paper, we describe a numerical method to compute the flow of liquid in an idealized diaper. The performance of this diaper is subsequently improved by optimizing the spatial distribution of the materials. The target is to maximize the volume of absorbed liquid. The assignment of material is at the level of the mesh discretization, permitting nearly arbitrarily complex designs. This approach, called topology optimization, belongs together with shape optimization to the field of structural optimization. The underlying optimization model is the so-called solid isotropic material with penalization (SIMP), dating back to Bends{\o}e \cite{Bendsoe:1989:OptimalShape}. Comparisons of its performance with alternative structural optimization approaches can be found in the literature \cite{Bendsoe:2003:Book,SigmundMaute:2013:Review}.

This paper is structured as follows. The model and the numerical method for the solution of the diaper problem are presented in \secref{sec:model} and \secref{sec:numsol}, respectively. In \secref{sec:topology_optimization} we introduce the topology optimization approach, including the formulation of the optimization problem. Our optimization results are presented in \secref{sec:results_bi} and \secref{sec:results_three}, with variants in \secref{sec:variants}. The conclusions are given in \secref{sec:conclusion}. 

\section{Simplified numerical model of a diaper}\label{sec:model}

\begin{figure}
\centering
\includegraphics[width=0.6\linewidth]{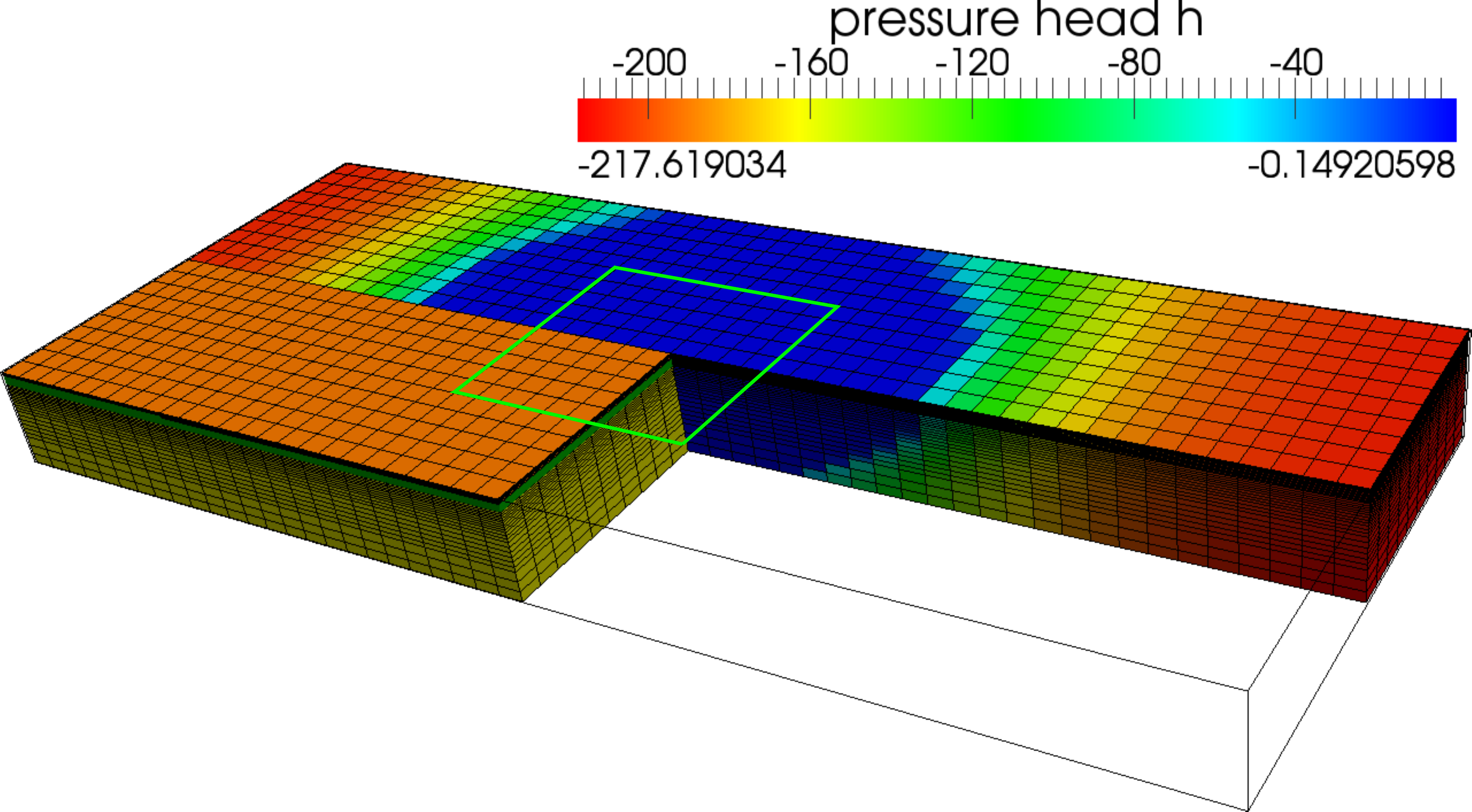}
\caption{\label{fig:domain} Computational domain of our diaper model. In the front part we show the three material layers of the standard layered design. The layers of material $A$ (orange) and $B$ (green) are thin, the thickest layer consists of material $C$ (yellow). The back part visualizes the state variable pressure head $h$ (in cm) at the end of the equilibration phase after a liquid discharge. The green frame indicates the boundary of the inlet patch where the liquid is discharged. 
}
\end{figure}
We model the diaper as a Cartesian box $\Omega$=$[0,L_x]\times[0,L_y]\times[-L_z,0]$ and gravity acts in the negative $z$ direction. The length, width and depth of the box are $L_x$=25\,cm, $L_y$=10\,cm and  $L_z$=2\,cm, respectively, and the total diaper volume $v=500\,\text{cm}^3$. All materials are considered as non-swelling porous media, so our model represents a diaper swollen to its maximum volume even if no water is present.  With this simplification a dry diaper can be seen as a porous medium of very high porosity, which models SAPs having the capacity to absorb a high volume of liquid. As the liquid is discharged, the pores get progressively filled. 

A schematic of the computational model of the diaper is shown in \figref{fig:domain}.  Liquid enters the box from the top at a 5\,cm $\times$ 5\,cm rectangle during a defined discharge time followed by an equilibration phase. The goal of the optimization problem is to propose a diaper design that can absorb more liquid than the reference three-layer design. The new design should use the same volume of material $A$, $B$ and $C$ but achieve a higher efficiency by optimizing their spatial distribution. The precise mathematical formulation of the optimization problem is given in \secref{sec:topology_optimization}.  

\subsection{Governing equations and material parameters}

\begin{figure}
\centering
\includegraphics[width=0.31\linewidth]{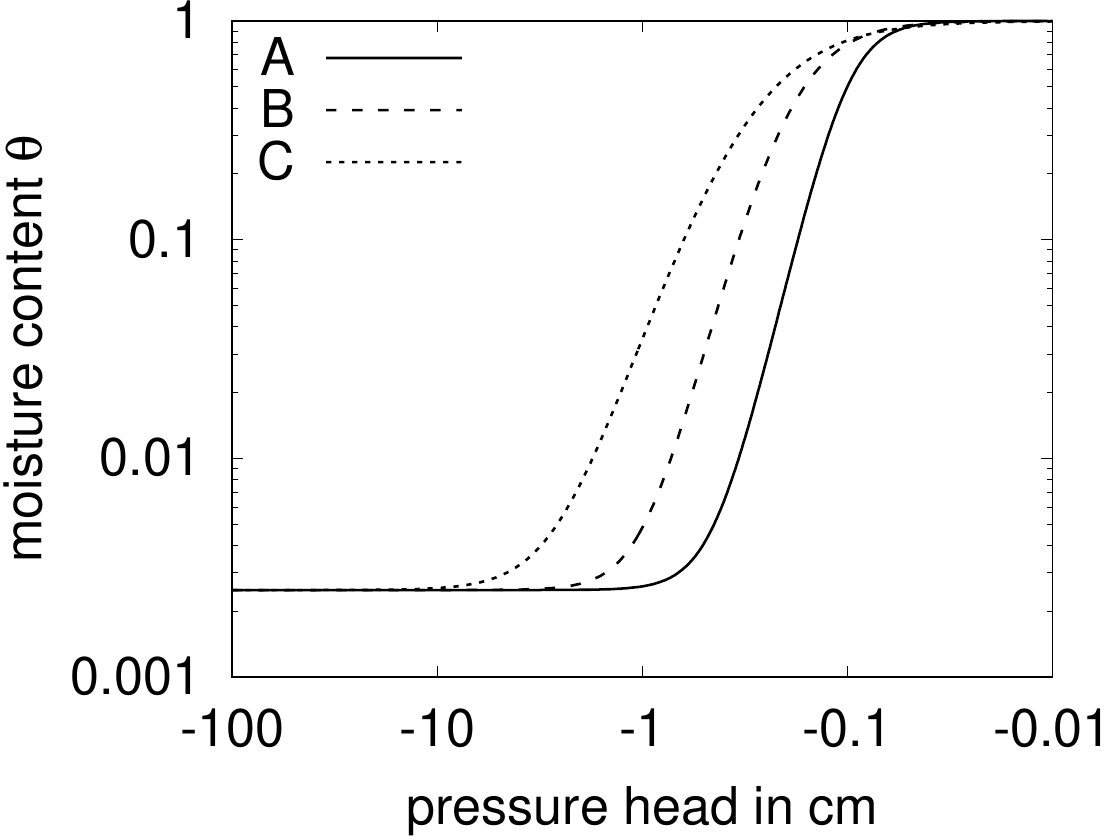} \; 
\includegraphics[width=0.31\linewidth]{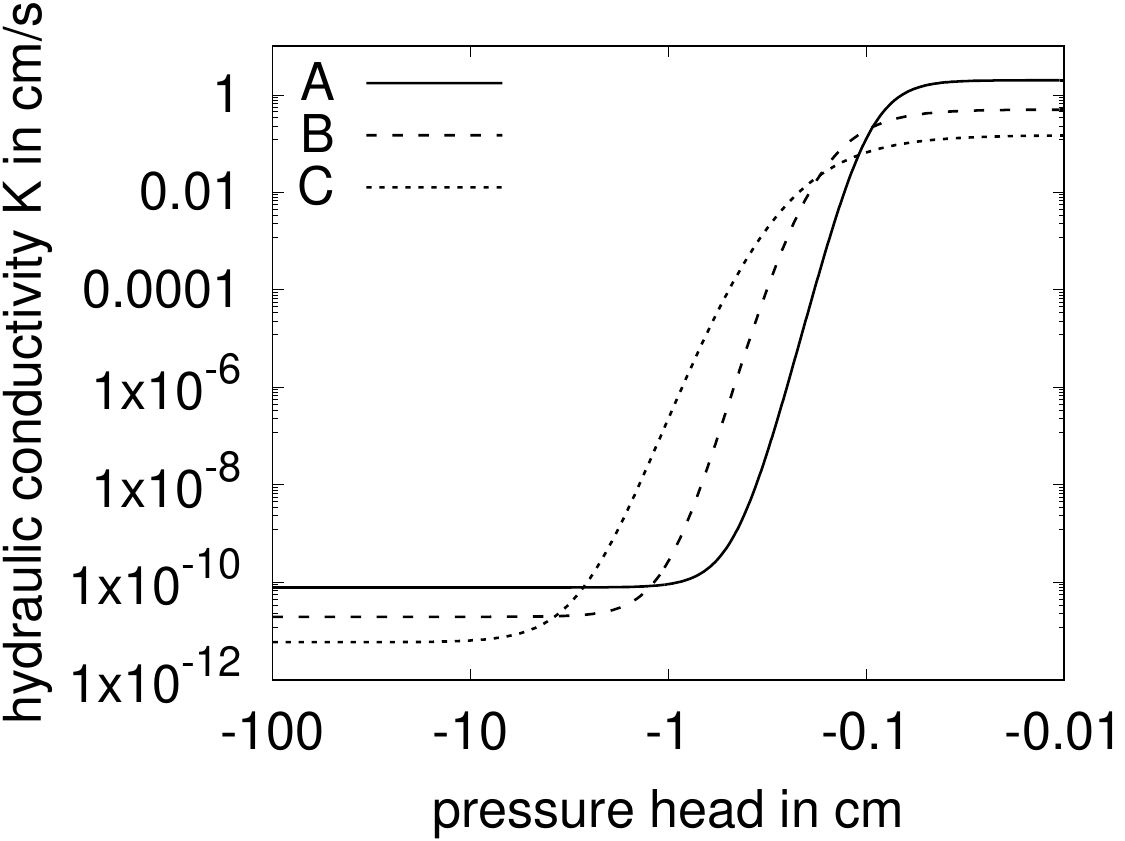} \;
\includegraphics[width=0.31\linewidth]{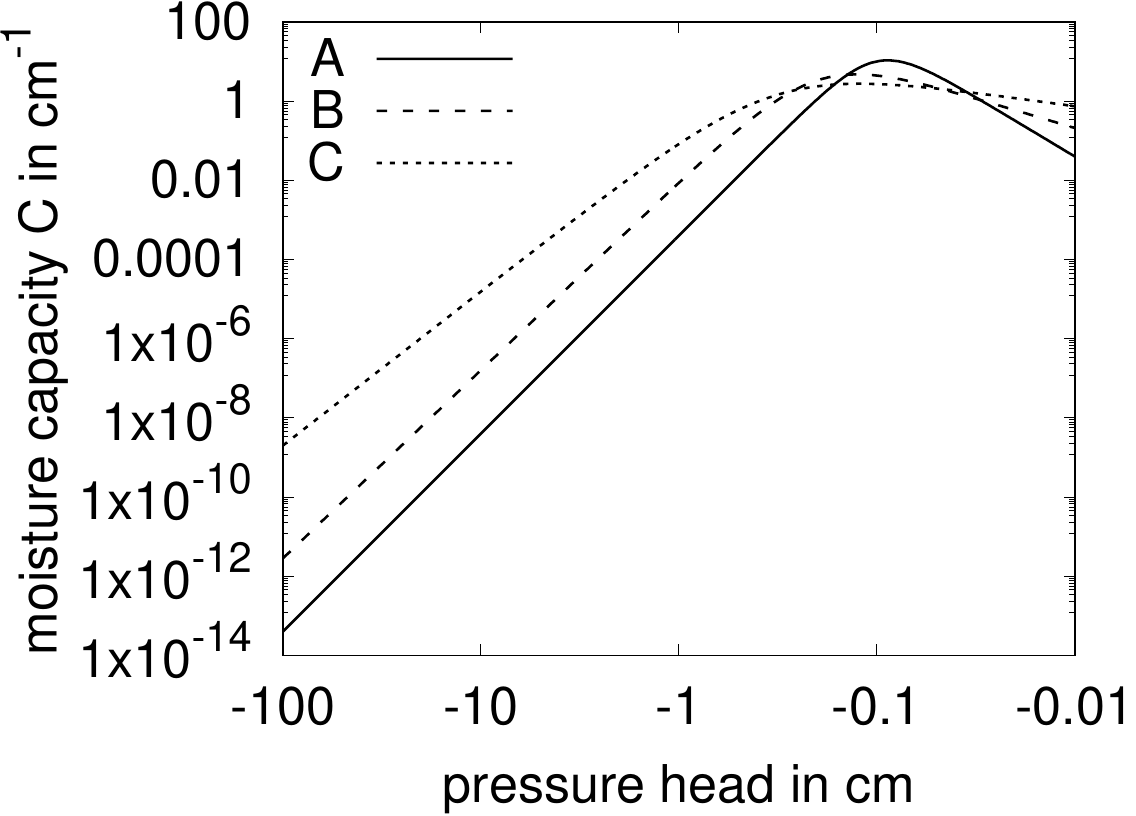} 
\caption{Properties of the diaper materials $A$, $B$ and $C$ as a function of the hydraulic pressure head. From left to right: moisture content, hydraulic conductivity and moisture capacity.\label{fig:matProp}}
\end{figure}

Richards' equation \cite{Richards1931} describes the motion of liquids in unsaturated porous media. In our formulation, the hydraulic pressure head $h=p/(\rho\,g)$ (in cm)  is used as the main variable, where $p$ is the capillary pressure, $\rho$=1g/cm$^3$ the liquid density (water at 20$^\circ$C) and $g$=981cm/s$^2$ the gravitational acceleration. A full liquid saturation corresponds to zero capillary pressure, $h=0$, whereas completely dry material has $h\rightarrow -\infty$. Richards' equation for the hydraulic pressure head reads
\begin{equation}
  \phi\frac{\partial \theta}{\partial t}=\nabla \cdot \left[K\nabla (h + z)\right],
  \label{eq:mRichards}
\end{equation}
where $\phi$ is the porosity of the material and $K$ the hydraulic conductivity. For a  solid material (without pores) $\phi$=0, whereas $\phi=1$  corresponds to the limit in which there is no solid matrix. The dimensionless moisture content $\theta(h)$ can in principle take values between $0$ and $1$, corresponding to dry and completely filled pores respectively. Here the van Genuchten parametrization is used to model the behavior of the SAPs
\begin{equation}
  \frac{\theta-\theta_{r}}{1-\theta_{r}}=\frac{1}{[1+(A_{h}h)^{n}]^{m}}.
  \label{eq:van_Genuchten}
\end{equation}
In practice a small residual moisture content $\theta_{r}$=0.0025 is used in order to avoid that the pressure head diverges ($h\rightarrow -\infty$) in dry material. The values of the material parameters $A_h$, $n$ and $m$ have been chosen to qualitatively model the absorption behavior of diapers, while keeping the problem tractable numerically (see \tabref{tab:ConversionParameter} for their specification). The values used for material C are similar to \cite{Diersch2011}, who compared their results to laboratory experiments with diapers. The left panel of \figref{fig:matProp} shows the moisture content as a function of the pressure head for the three materials. Note how the uppermost material $A$ remains essentially dry up to $h$=-1\,cm, whereas bottommost material $C$ starts to get significantly wet already at $h$=-10\,cm. The buffer material $B$ features intermediate properties. 
\begin{table}
  \centering
  \begin{tabular}{c|c|c|c|l}
 & $A$ & $B$ & $C$ & {Description}\\
\hline
$\phi$ & 0.9 & 0.9 & 0.95 & {porosity}\\
\hline
$A_h$ in 1/cm & 10 & 5 & 3 & {coefficient in \eqnref{eq:van_Genuchten}}\\
\hline
n & 4 & 2.5 & 1.7 & {coefficient in \eqnref{eq:van_Genuchten}}\\
\hline
m & 1.0 & 1.5 & 1.7 & {coefficient in \eqnref{eq:van_Genuchten}}\\
\hline
K$_0$ in cm/s & 2.0 & 0.5 & 0.15 & {maximum conductivity}
  \end{tabular}
\caption{Parameters defining the properties of the diaper materials $A$, $B$ and $C$ (see \eqnref{eq:van_Genuchten}--\eqnref{eq:C_h} ).}\label{tab:ConversionParameter}
\label{tab:material}
\end{table}

The materials are assumed isotropic. The hydraulic conductivity is a scalar
\begin{equation}
  K=\frac{k\rho g}{\mu},\label{eq:K_k}
\end{equation}
where $k$ is the permeability of the material and $\mu$=0.01g/(cm\,s) the dynamic viscosity of water at 20$^\circ$C. As materials are progressively wetted, their hydraulic conductivity increases, and this is modeled by the expression
\begin{equation}
  K(\theta)=K_{0}\theta^4. \label{eq:K_theta}
\end{equation} 
The values of the maximum conductivity $K_0$ for each material can be found in \tabref{tab:ConversionParameter}. It is worth noting that the maximum conductivity of material $A$ is significantly larger than that of $C$, so that liquid is rapidly evacuated towards the bottom of the diaper. The mid panel of \figref{fig:matProp} shows the conductivity as a function of pressure head for the three materials.  

Another key material property is the moisture capacity 
\begin{equation}
  C(h)=\total{\theta}{h},\label{eq:C_h}
\end{equation}
which quantifies how much liquid can be stored at a given pressure head. The three materials feature peak moisture capacity at different pressures. Material $A$ absorbs liquid efficiently only when it is quite wet, i.e. $h$=-0.1\,cm, whereas  material $C$ has a quite flat profile, allowing liquid to be efficiently absorbed for a wide range of pressure heads. Material $B$ features an intermediate behavior. 

We note that equation \eqref{eq:mRichards} is known as the mixed form of Richards' equation. By using the relationship between moisture content and pressured head $\theta(h)$, one obtains the head form of Richards' equation
\begin{equation}
  \phi C(h) \frac{\partial h}{\partial t}=\nabla \cdot \left[K\nabla (h + z)\right],
  \label{eq:hRichards}
\end{equation}
which has the advantage that $h$ is the only unknown. Hence Richards' equation is essentially a transient Darcy equation for the pressure head $h$, where the effective porosity $\phi C(h)$ and the hydraulic conductivity $K(h)$ depend nonlinearly on $h$. Because a direct discretization of the head form of the equation results in large mass losses \cite{Celia1990}, Richards' equation was here solved in its mixed form \eqref{eq:mRichards} as detailed in \secref{sec:numsol}.

\subsection{Initial and boundary conditions}

At time $t$=0 the diaper is dry and the liquid discharge phase begins. When liquid is deposited on top of a porous medium, such as the diaper considered here, it forms a puddle, which is then subsequently absorbed. The absorption speed and hence the flux at the surface are unknown a priory. In fact, the flux depends nonlinearly on the permeability and capacity, which depend on the moisture content itself. In a nutshell, the  flux is a function of the material properties and of the liquid distribution inside the diaper. We model the puddle as a 5\,cm $\times$ 5\,cm rectangle centered at $(12.5,5,0)$, where fully wet material ($\theta=1$) is imposed as boundary condition. At the rest of the domain's boundary the flux is set to zero to enforce that no liquid exits the diaper. During a subsequent equilibration  phase, zero flux boundary conditions are imposed at all boundaries. In this problem, the wetting dynamics and the content of water which can be absorbed is mainly determined by the large moisture gradients near the inlet, and the large gradients of permeability at the interface between materials. This requires very fine meshing in the axial directions and small time-step sizes during the liquid discharges. 

\subsection{Simulation of the three-dimensional reference design}\label{sec:reference}

\begin{figure}[!th]
\centering
\includegraphics[width=0.44\linewidth]{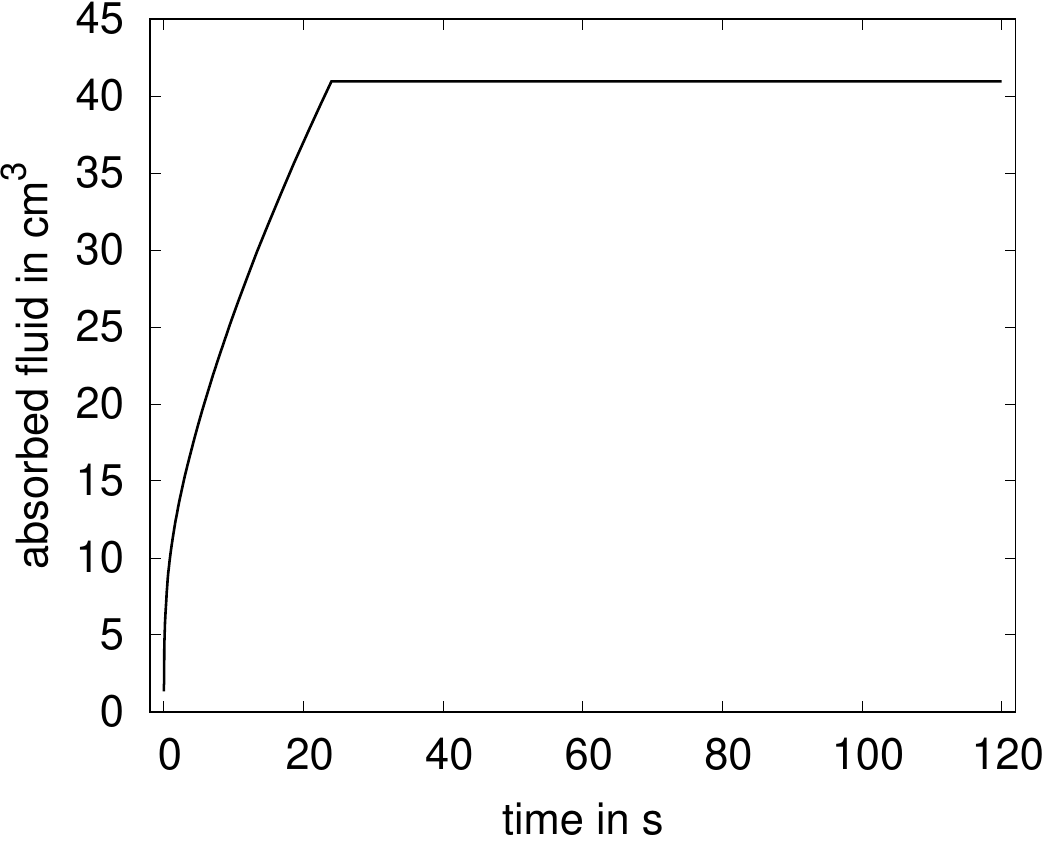}
\includegraphics[width=0.52\linewidth]{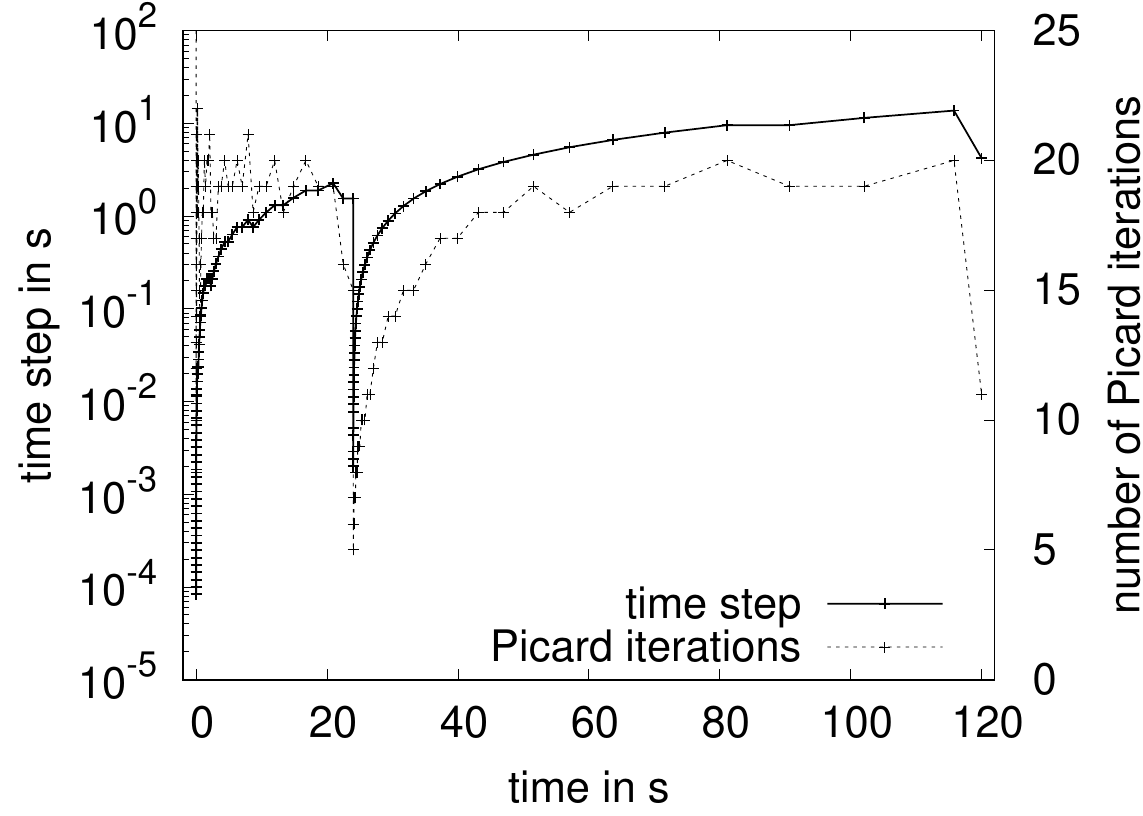}  
\caption{\label{fig:sim_std} Numerical simulation of the reference three-layer design. Left panel: during a 24\,s discharge phase the total liquid content increases. Right panel: the adaptive time step algorithm targets at 20 Picard iterations per time step.}
\label{fig:num}
\end{figure}

We considered a  discharge resulting in a puddle of 24\,s duration, which is followed by an equilibration phase of 120\,s. The left panel of \figref{fig:num} shows the evolution of the total liquid content in the diaper with time. At $t=0$, there is a rapid increase of the liquid content, because of the strong pressure gradients between the wet inlet and dry diaper. As the liquid progressively fills the diaper, gradients are reduced and the absorption rate decreases. During the equilibration phase liquid is distributed throughout the diaper, while the total liquid content remains constant. This indicates that our simulation conserves mass. We note that because of the dominance of capillary forces, gravity was neglected in our simulations. This corresponds to eliminating the term $z$ from the right-hand-side of Richards' equation  \eqref{eq:mRichards} and is justified in the next section.
   
\subsection{Influence of gravity}

\begin{figure}[!ht]
 \centering
 {\includegraphics[width=.45\textwidth]{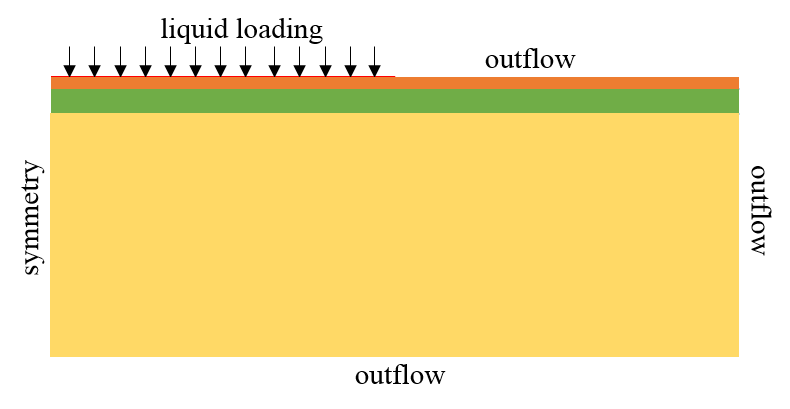} } 
  \caption{\label{fig:BCs}Compuational domain and boundary conditions of 2D three-layer simulation. The orange, green and yellow color indicate material $A$, $B$ and $C$, respectively.}
\end{figure}

\begin{figure}[!ht]
 \centering
 \subfloat[three layer simulation with gravity ]
 {\includegraphics[width=.4\textwidth]{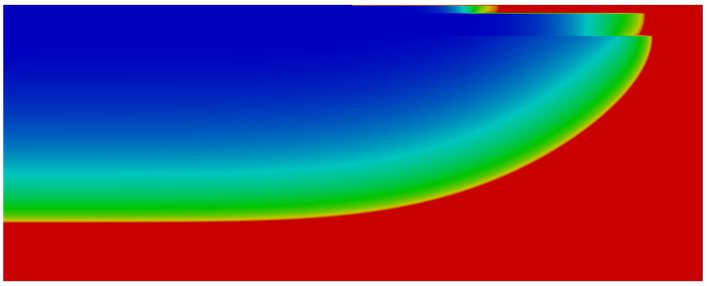} \label{fig:gravity}}
 \subfloat[three layer simulation without gravity ] 
 {\includegraphics[width=.4\textwidth]{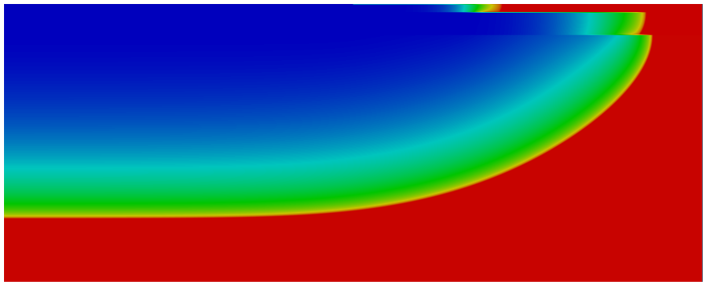} \label{fig:nogravity}} 
  \subfloat[ ] 
 {\includegraphics[width=.063\textwidth]{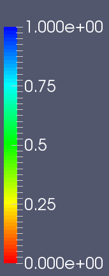} \label{fig:ColorBar}} 
  \caption{\label{fig:gravityComp} Moisture of a two-dimensional simulation of the reference three-layer design at $t=$5s. (a) with gravity. (b) without gravity. (c) scale of the moisture colormap, where dry (wet) material is shown as red (blue).}
\end{figure}

\begin{figure}[!ht]
 \centering
 {\includegraphics[width=.35\textwidth]{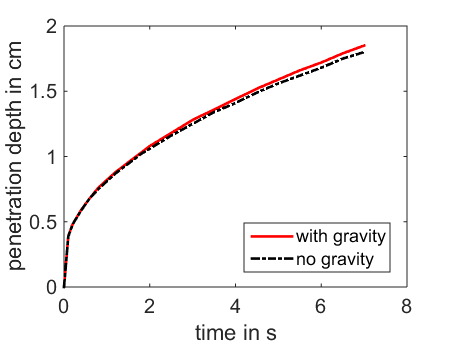} } 
  \caption{\label{fig:penetration} Comparison of deepest penetration location with gravity and without gravity as a function of time for the simulation shown in \figref{fig:gravityComp}.}
\end{figure}

In order to study the influence of gravity on the wetting dynamics, a two-dimensional three-layer simulation of the reference structure was performed. We considered the 5\,cm $\times$ 2\,cm symmetric slice cut across the width and depth shown in \figref{fig:domain} and the boundary conditions as given in \figref{fig:BCs}. The liquid load was applied on the left half on top of the domain, the right half of the top boundary, as well as the right and bottom sides, are set to outflow to let the liquid pass through, the left boundary is set to be symmetric in order to reduce the computing time. The simulation started with the liquid coming from the loading region and ran until liquid started to pass through the right boundary. The whole simulation time was 7.5s.
\figref{fig:gravityComp} shows that the moisture content of the reference three-layer design at $t=$5s is indistinguishable in the cases with and without gravity. \figref{fig:penetration} shows the development of the deepest location of penetration with respect to time. It can be observed that although the deepest location is slightly larger with the effects of gravity, the difference is still fairly small. Furthermore, the simulation with gravity is substantially more expensive because it requires much higher grid resolution, resulting in too expensive simulations for three-dimensional optimization. In what follows, gravity is neglected. 


\section{Numerical method}\label{sec:numsol}

Because of the nonlinearities in Richards' equation, care needs to be taken to enforce mass conservation in the numerical simulations. Celia~\emph{et al.}~\cite{Celia1990} developed a modified Picard iteration of the mixed form of the equations~\eqref{eq:mRichards}, which conserves mass in free drainage problems. Although several improvements and extensions of this scheme have been proposed in the literature~\cite{Huang1996,Zhang2005,McBride2007,Phoon2007,Zeng2008}, we here employed the popular method of Celia~\emph{et al.}~\cite{Celia1990} because it gives an excellent compromise between accuracy and efficiency in simple geometries, and simple implementation. The reader is referred to List \emph{et al}.~\cite{list2016study} for a recent review on methods to linearize the Richards' equation. In what follows, we briefly summarize the modified Picard scheme used in this paper. Time was advanced with the implicit Euler scheme, 
\begin{equation}
\phi\frac{\theta^{i+1}-\theta^{i}}{\Delta t}=\nabla\cdot\left[K^{i+1}\left(\nabla h^{i+1}+z\right)\right]\label{eq:Euler}
\end{equation}
which required the solution of a nonlinear problem every time step $i+1$. This was solved iteratively with the Picard method~\cite{Celia1990}, whereby the moisture content at time step $i+1$ is expanded in a Taylor series with respect to $h$ as
\begin{equation}
\theta^{i+1,j+1}=\theta^{i+1,j}+C^{i+1,j}\left[h^{i+1,j+1}-h^{i+1,j}\right]+O(h.o.t),
\end{equation}
and the superscript $j+1$ denotes the current Picard iteration. Neglecting high order terms and inserting this expression into \eqref{eq:mRichards}, the Picard iteration reads
\begin{equation}
\phi C^{i+1,j}\frac{h^{i+1,j+1}-h^{i+1,j}}{\text{\ensuremath{\Delta}t}}=\nabla\cdot\left[K^{i+1,j}\nabla (h^{i+1,j+1}+z)\right]-\phi\frac{\theta^{i+1,j}-\theta^{i}}{\Delta t}.\label{eq:PicardItr}
\end{equation}
Hence at each Picard iteration a (linear) Helmholtz equation for $h^{i+1,j+1}$ must be solved. The spatial derivatives were discretized with the second-order central finite-volume method. Our code was implemented in OpenFOAM~\cite{OpenFOAM} and was successfully benchmarked against~\cite{Celia1990}. Note how eq.~\eqref{eq:PicardItr} is identical to the Picard equation for the head form of the Richards' equation, but with the additional term in the right-hand-side containing the moisture content.

An important difficulty in our problem is that the diaper contains three materials with very different properties. Numerical studies of two-layer~\cite{Cheng2003,Kalinka2011} and multi-layer~\cite{Heise2006} domains have been reported in the soil-science literature. Note however that the material properties of diapers varies significantly and are numerically challenging to treat. 
Our analysis domain has a discretization of $50 \times 20 \times 35$ cells, where the cell height in $z$-direction grows from very thin cells at the inlet plane towards rather thick ones at the other end.

A further difficulty in our problem is that the time-step required for the Picard iterations to converge varies across several orders of magnitude. At the beginning of the liquid discharge, small time-step sizes ($\Delta\, t\approx10^{-4}$s) were required, whereas during the equilibration periods $\Delta t\approx1$s. Hence the time-step size was dynamically adjusted based on the number of Picard iterations, whereby 20 iterations were considered as optimal. Then depending on whether 20 Picard iterations were exceeded or not in the previous time step, $\Delta t$ was divided or multiplied by 1.1. If the number of Picard iterations was below 6, the next time step was increased by 1.5. In addition to this criterion to determine the time-step dynamically, we set a hard limit of 50 Picard iterations. If this was exceeded,  $\Delta t$ was reduced by $75\%$ and the time-step was re-computed. If necessary, this was repeated until the number of iterations dropped below 50. Finally, we note that $\Delta t$ was adjusted to resolve the switching between inflow and zero flux boundary conditions at $t=24$s. This was done to avoid wasting time by reaching the 50 iteration limit repeatedly. The time-step size and number of Picard iterations required in the simulation of the reference three-layer design is shown in the right panel of \figref{fig:num}.

\subsection{Treatment of the multi-material domain}\label{sec:multimaterial}

Our solver was implemented to handle three materials in one function. For this purpose, we defined a three-element vector determining the material of each computational volume $j$, 
\begin{equation}
\bw_j = (w^{A}_j, w^{B}_j, w^{C}_j) \in \{0,1\}^3,
\label{eqn:w_j_discrete}
\end{equation}
where
\begin{equation}
|| \bw_j || = w^{A}_j + w^{B}_j + w^{C}_j = 1,
\label{eqn:w_j_norm}
\end{equation}
and only one component of $\bw_j$ was set to one and the others to zero at each computational volume, corresponding to an acceptable (discrete) material choice. The generalized material properties can be written as follows
\begin{eqnarray}
  \theta_j(\bw_j, h)&=&w^{A}_j\theta^{A}_j(h)+w^{B}_j\theta^{B}_j(h)+w^{C}_j\theta^{C}_j(h),\\  \label{eqn:theta_by_w_j}
  C_j(\bw_j, h)&=&w^{A}_jC^{A}_j(h)+w^{B}_jC^{B}_j(h)+w^{C}_jC^{C}_j(h),\\  \label{eqn:C_by_w_j}
  K_j(\bw_j, \theta)&=&w^{A}_jK^{A}_j(\theta)+w^{B}_jK^{B}_j(\theta)+w^{C}_jK^{C}_j(\theta),\\  \label{eqn:K_by_w_j}
  \phi_j(\bw_j)&=&w^{A}_j\phi^{A}_j+w^{B}_j\phi^{B}_j+w^{C}_j\phi^{C}_j.  \label{eqn:phi_by_w_j}
\end{eqnarray}

\section{Topology Optimization}\label{sec:topology_optimization}

We applied topology optimization to distribute the materials within the design domain $\Omega$. In the $x$- and $y$-directions, the design discretization of $\Omega$ was identical to the finite-volume discretization used for the numerical simulation. In the $z$-direction, each design cell could contain several finite volumes. In the following, we refer to $N$ as the total number of design cells (with index $e$), whereas $D$ is the total number of computational volumes (with index $j$).  

\subsection{Parametrization}
\label{sec:parametrization}

To perform gradient based optimization, we relaxed the selection variable $\bw_j$ \eqnref{eqn:w_j_discrete} to a continuous weighting variable \begin{equation}
  \brho_e \in [0, 1] \times [0, 1] \times [0, 1] = [0, 1]^3 \subset \mathbb{R}^3
  \label{eqn:rho_j_threemat}
\end{equation}
with $1 \leq e \leq N$ and the constraint
\begin{equation}
  \rho^A_e + \rho^B_e + \rho^C_e = 1.
  \label{eqn:rho_j_L1_norm}
\end{equation}
In topology optimization, the design variables $\brho_e$ are commonly called \textit{pseudo density}. We use bold symbols and lowercase letters to indicate vectors. Hence the full vector of design variables of size $3\,N$ is given in bold as is the set of variables assigned to a single element for the multimaterial case. Three scalar variables per element in \eqnref{eqn:rho_j_L1_norm} are written as elementwise vector in \eqnref{eqn:rho_j_threemat} which spans the three-dimensional continuous space beween 0 and 1. In standard topology optimization, the distribution of a single material is optimized, whereas we here aim to solve a multi-material optimization problem similar to the class of problems treated by Erik Lund and co-workers using the discrete material optimization approach, see e.g. \cite{Hvejsel:2011} and earlier work. To enforce condition \eqnref{eqn:rho_j_L1_norm}, $N$ constraints are required when we model the optimization problem. This approach can be easily extended to more than three materials.

When we restrict ourselves to only two materials, a scalar design variable $\rho_e$ is sufficient with
\begin{equation}
\rho_e \in [0, 1]
\label{eqn:rho_e_scalar}
\end{equation}
and the weights of two arbitrary materials, e.g. $A$ and $C$, are $\rho_e^A=\rho_e$ and $\rho_e^C=1-\rho_e$ with $\rho_e^B=0$. This approach is known as bi-material optimization (see \cite{Sigmund:96:Composites}) and the norm condition is implicitly fulfilled.
    
The idea of the optimization is to construct material properties \eqnref{eqn:theta_by_w_j}--\eqnref{eqn:phi_by_w_j} using $\brho_j$ instead of $\bw_j$. We note that only discrete material selections $\brho_j \in \{0, 1\}^3$ allow a meaningful physical interpretation. 
With the final designs sufficiently close to discrete material distributions the effect of mixed material on the different material properties \eqnref{eqn:theta_by_w_j}--\eqnref{eqn:phi_by_w_j} plays only a minor role and is thus not needed to be analyzed it in detail.  
  
\subsection{Sensitivity Analysis}\label{sec:sensitivity_analysis}  

The purpose of the sensitivity analysis is to provide the first-order derivatives of the state dependent functions with respect to the design variables, entering the optimization problem as objective function or constraints. Having the sensitivities available, a suitable first-order optimizer toolbox can be used to solve the optimization problem. In general, this is vastly more efficient than utilizing derivative-free optimizers \cite{Sigmund:2011:NonGradientBashing} and is here of particular importance, as we solve a high-dimensional nonlinear problem (with $3\,N$ design variables for the three-material case).

\subsubsection{Semi-Discrete System} 

We consider the head form of Richards' equation \eqnref{eq:mRichards} and denote explicitly the dependencies on the state variable pressure head $h$ and the material parametrization by the design vector $\brho$. If gravity is neglected, this reads as
\begin{equation}
\phi (\rho)\, C(\rho, h) \dt{h} - \nabla \cdot K(\rho, h) \nabla h = 0,
\label{eqn:richards_h_based}
\end{equation} 
and after spatial discretization with the finite-volume method, we obtain the semi-discrete system of equations
\begin{equation}
\bM(\brho, \bh)\, \dt{\bh} - \bK(\brho, \bh)\,\bh = \bzero.
\label{eqn:algebric_semi_discrete}
\end{equation}
The system is still continuous in time. The mass matrix $\bM$ and stiffness matrix $\bK$ are discretizations of their continuous counterparts given in \eqnref{eqn:richards_h_based}, see \cite{Celia1990}. 
  
We can now formulate a semi-discrete algebraic generic optimization problem with objective function $J$ depending on the design $\brho$ and state $\bh$, as well as a constraint function $g(\brho)$ independent of the state. The number of constraint functions and their actual type (equality or inequality) is not of importance for the following sensitivity analysis       
\begin{eqnarray}
\label{eqn:generic_problem}
\min_\brho  J(\brho, \bh) \Big \vert_{t=T} \nonumber &&\\
\text{s.t. } \bM(\brho, \bh)\,\dt{\bh} - \bK(\brho, \bh)\,\bh  &=& \bzero, \\
g(\brho) &=& 0. \nonumber
\end{eqnarray}
Note that we are interested in the function value for $J$ at final time $T$ only.   

\subsubsection{Derivation of the Gradients}

The sensitivity analysis presented in the following closely follows Dahl, Jensen and Sigmund \cite{Dahl:2008:Transient}, however our system \eqnref{eqn:algebric_semi_discrete} is nonlinear and has no second time-derivative. To find the adjoint formulation, we consider the residual of the state equation point-wise
\begin{equation}
\tM \big \vert_{t} \,\dt{\bh(t)} - \tK \big \vert_{t}\,\bh(t) = \bzero \quad \forall t \in [0, T],
\label{eqn:algebraic_time}
\end{equation}
where $\tM\big \vert_{t}= \bM(\brho, \bbh) \big \vert_{t}$, $\tK \big \vert_{t}= \bK(\brho, \bbh) \big \vert_{t}$
and $\bbh: [0, T] \to \mathbb{R}^D$ is the converged state solution of \eqnref{eqn:algebric_semi_discrete}, where $D$ is the number of computational volumes resulting from the space discretization. The tilde denotes explicit dependence on $\brho$.

Introducing a Lagrange multiplier $\blmbd: [0, T] \to \mathbb{R}^D$ we obtain 
\begin{equation*}
  \Phi = \Phi(\brho, \bh) \Big \vert_{t=T} = J(\brho, \bh) \Big \vert_{t=T} + \int_0^{T} \blmbd^\top \left(\tM \,\dt{\bh} - \tK\,\bh \right)\, \intd t,
\end{equation*} 
where the integral term in the equation is zero because of the residual \eqnref{eqn:algebraic_time} and the objective function value is not changed. Differentiation yields
\begin{eqnarray*}
\total{\Phi}{\rho_e} & = &\drho{J} \Big \vert_{t=T} + \sens{J}{h} \, \drho{\bh}
\Big \vert_{t=T} \\
& + & \int_0^{T} \blmbd^\top \tM  \frac{\partial^2\bh}{\partial t\, \partial \rho_e} \,\intd t - \int_0^{T} \blmbd^\top \tK \drho{\bh} \,\intd t \\
& + & \int_0^{T} \blmbd^\top \drho{\tM} \,\dt{\bh} \,\intd t
- \int_0^{T} \blmbd^\top \drho{\tK}  \,\bh \,\intd t.
\end{eqnarray*}
The critical term to determine is the derivative of the state $\bh$ with respect to the design variables $\rho_e$. In the following, we show that this vanishes with the help of the Lagrange multiplier. Applying integration by parts yields
\begin{eqnarray*}
\total{\Phi}{\rho_e} & = &\drho{J} \Big \vert_{t=T} + \sens{J}{h} \, \drho{\bh}
\Big \vert_{t=T} +  \blmbd^\top \tM \drho{\bh} \bigg \vert_{t=T} -  \blmbd^\top \tM
\drho{\bh} \bigg \vert_{t=0} \\ 
& - & \int_0^{T} \dt{\blmbd}^\top \tM \drho{\bh} \,\intd t -
\int_0^{T} \blmbd^\top \tK  \drho{\bh} \,\intd t \\
& + & \int_0^{T}  \left( \blmbd^\top \drho{\tM} \,\dt{\bh}
- \blmbd^\top \drho{\tK}  \,\bh \right) \,\intd
t.
\end{eqnarray*}
By collecting the terms with $\drho{\bh}$ we obtain
\begin{eqnarray*}
\total{\Phi}{\rho_e} & = &\drho{J} \Big \vert_{t=T}  + \int_0^{T}  \left( \blmbd^\top \drho{\tM} \,\dt{\bh}
- \blmbd^\top \drho{\tK}  \,\bh \right) \,\intd
t \\
& - & \int_0^{T} \left( \dt{\blmbd}^\top \tM  + \blmbd^\top \tK  \right) \drho{\bh} \,\intd t \\
& + & \left( \blmbd^\top \tM + \sens{J}{h} \right) \drho{\bh} \bigg
\vert_{t=T} - \blmbd^\top
\tM \drho{\bh} \bigg \vert_{t=0}. 
\end{eqnarray*}
The unknown terms $\drho{\bh}$ vanish if $\blmbd$ solves the terminal value adjoint equation
\begin{equation*}
\tM \dt{\blmbd} +  \tK \, \blmbd = \bzero
\end{equation*}
and
\begin{equation*}
\tM \, \blmbd \bigg \vert_{t=T} = - \left(\sens{J}{h}\right)^\top \bigg
\vert_{t=T}.
\end{equation*}
Here we make use of the fact that $\tM$ and $\tK$ are symmetric matrices. As there is no dependency of the initial state solution on the initial material distribution (\cite{Dahl:2008:Transient}), also the term for $t=0$ vanishes.

The terminal value adjoint problem is transformed into an initial value problem by variable substitution. Hence the gradient is calculated as
\begin{equation}
\total{\Phi}{\rho_e} = \drho{J} \Big \vert_{t=T}  + \int_0^{T}  \left( \blmbd^\top \drho{\tM} \, {\bh}
- \blmbd^\top \drho{\tK}  \,\bh \right) \,\intd t,
\label{eqn:dt_drho} 
\end{equation}  
where $\blmbd$ solves the initial value adjoint problem
\begin{equation}
-\tM \dt{\overline{\blmbd}} +  \tK \, \overline{\blmbd} = \bzero \; \Bigg
\vert_{T-\tau},\quad \overline{\blmbd}(\tau) = \blmbd(T-t),
\label{eqn:adjoint}
\end{equation}
with initial condition
\begin{equation}
\tM \, \overline{\blmbd} \bigg \vert_{\tau=0} = - \left(\sens{J}{h}\right)^\top
\bigg \vert_{t=T}.
\label{eqn:adjoint_initial}
\end{equation}
Note the change of sign for the time derivative in \eqnref{eqn:adjoint}. In the adjoint problem the mass and stiffness matrices do not depend on $\overline{\blmbd}$, hence no Picard iterations are necessary for the numerical solution of the adjoint problem. On the other hand, we need to store for each time step the converged state $\bh$ and store or reconstruct $\tM$ and $\tK$ as well. 

Because of \eqnref{eqn:adjoint_initial}, we need to solve a separate transient adjoint problem for any state dependent function.  The negative derivative of the function with respect to the state is the initial condition.
    
\subsection{Cost and Constraint Functions}\label{sec:functions}

The total volume of absorbed liquid is
\begin{equation}
J(\brho, \bh) = \sum_{j=1}^D \nu_{e(j)} \phi_j(\rho_{e(j)}) \,\theta_j\big(\rho_{e(j)}, \bh)\big),
\label{eqn:f_total}
\end{equation}
where $\nu_{e(j)}$ is the volume of the computational cells $\Omega_j$ and $e: \mathbb{R}^D \to \mathbb{R}^N$ maps design cells to computational cells. Again, we skip explicit dependence on time in the notation. For the objective function, we are interested in the function value only at final time $t=T$. We recall that the reason why we distinguish between design and computational cells is that we must use a very fine non-uniform mesh discretization in $z$-direction for numerical  reasons. We thus combine several computational cells to one design cell.
 
The amount of a given material, also known as resource constraint, is 
\begin{equation}
v(\rho) = \frac{1}{|\Omega|} \sum_{e=1}^N v_e\, \rho_e,
\label{eqn:vol-constraint}
\end{equation}
where $v_e$ is the volume of design cell $e$. For the bi-material case, $v(\brho)$ simply measures the amount of material $A$. 
The amount of the second material is implicitly given by $1-v(\brho)$. For the three-material case we require functions for two materials, here we choose the materials $A$ and $B$ such that the volume constraints $v^A(\brho)$ and $v^C(\brho)$ add to $\rho_e^A$ and $\rho_e^C$, respectively.     

In many applications in topology optimization, it is necessary to regularize the design space~\cite{Sigmund:98:NumericalInstabilities}. Typically the change of the design from cell to cell is limited by filtering the design \cite{Bruns:01:NonLin} or with the Heaviside projection method \cite{Guest:04:MinLengthScale}. Here we chose to use explicit regularization by slope constraints~\cite{Petersson:98:Slope}. The principal idea of slope constraints is to restrict the partial derivative of the design 
\begin{equation}
  s(\brho) = \left| \frac{\partial \rho}{\partial x_i} \right| \leq c_s, \quad i=1,2,3,
  \label{eqn:slope} 
\end{equation}
which are implemented with forward finite differences for every design element
\begin{eqnarray}
  | \rho_{i\,j\,k} - \rho_{i+1\,j\,k} | & \leq & c_s \quad \forall \; 1 \leq i \leq n_x-1, \; 1 \leq j \leq n_y,   \; 1 \leq k \leq n_z, \\
  | \rho_{i\,j\,k} - \rho_{i\,j+1\,k} | & \leq & c_s \quad \forall \; 1 \leq i \leq n_x,   \; 1 \leq j \leq n_y-1, \; 1 \leq k \leq n_z, \\
  | \rho_{i\,j\,k} - \rho_{i\,j\,k+1} | & \leq & c_s \quad \forall \; 1 \leq i \leq n_x,   \; 1 \leq j \leq n_y,   \; 1 \leq k \leq n_z-1. 
\end{eqnarray}
Choosing $c_s=1$ disables the slope constraints, $c_s=1/3$ requires at least two intermediate design cells for a full change from 0 to 1, or vice versa. Slope constraints come with a high number of linear constraints. Here we used the commercial SNOPT~\cite{Gill:02:SNOPT} package as optimizer to handle the high number of linear constraints. In our three-material case, it is sufficient to apply slope constraints on two materials. We want to mention that alternatively other forms of regularization, like density filtering, might be used.

Regularization still leaves room for isolated spots of intermediate material. One possible reason for this to happen might be that such material is irrelevant or negligible, e.g. in areas of the design domain too far away to be reached by moisture during the simulation. Moreover, surplus material stemming from the resource constraints might be distributed without benefit. 

As a remedy against mixed materials, we used constraints of the type
\begin{equation}
g(\brho) = \frac{1}{N} \sum_e^N 4\,\rho_e\,(1-\rho_e),
\label{eqn:grayness}
\end{equation}     
which are known as grayness constraints. Binary material (zero or one) has grayness zero, complete intermediate material (0.5) results in grayness one. Grayness constraints are known for the tendency to lock designs in local optima (which might be poor) as a smooth transition from void to solid or solid to void requires a temporary increase of the grayness which might violate the constraint.

A second problem is related to the concurrent application of slope constraints, which enforce intermediate densities at material interfaces. A too strict grayness constraint acts indirectly as control of the size of the interface, which might have a much more significant impact than mere removal of spots. For this reason, we applied grayness constraints only in a second stage of the optimization process starting from an already optimized design. In order to obtain a bound for the grayness constraint, we evaluated the grayness resulting from active slope constraints and used this value with a slight supplement.        

\subsection{Problem Formulation}
\label{sec:problem}

\begin{figure}[]
  \centering
  \includegraphics[width=.2\textwidth]{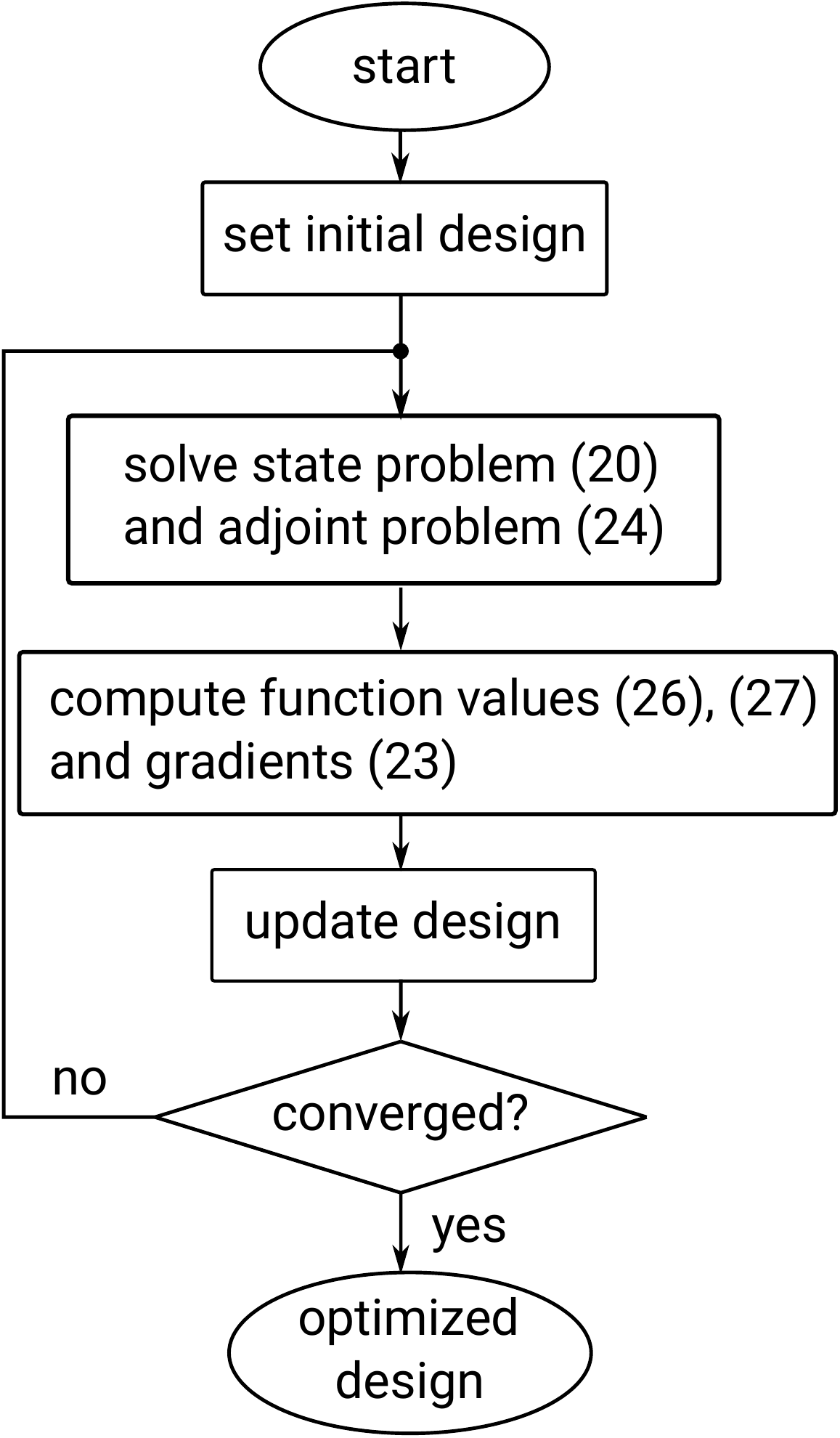}  \caption{\label{fig:flow_chart}General flow chart of the optimization problem. The referenced equations are given in brackets.}
\end{figure}

Having introduced the objective and constraints, we can now formulate the optimization problem, the general work flow is shown in \figref{fig:flow_chart}. For the bi-material case we optimize for the materials $A$ and $C$ with the design variable $\rho_e$ and $1-\rho_e$ expressing the fraction of material $A$ and $C$, respectively. Material $B$ has contribution zero. The bi-material optimization problem reads
\begin{eqnarray}
\max_\brho && J(\brho, \bh) \Big \vert_{t=T}  \label{eqn:prob_bimat_J} \\
\text{s.t. } \bM(\brho, \bh)\,\dt{\bh} - \bK(\brho, \bh)\,\bh  & = & \bzero, \label{eqn:prob_bimat_state}\\
v(\brho) & = & v_A^*, \label{eqn:prob_bimat_v}\\
g(\brho) & \leq & c_g, \label{eqn:prob_bimat_g} \\
 | \rho_{i\,j\,k} - \rho_{i+1\,j\,k} | & \leq & c_s \quad \forall \; i \leq n_x-1, \; j \leq n_y, \; k \leq n_z, \label{eqn:prob_bimat_s_1} \\
 | \rho_{i\,j\,k} - \rho_{i\,j+1\,k} | & \leq & c_s \quad \forall \; i \leq n_x,   \; j \leq n_y-1, \; k \leq n_z, \label{eqn:prob_bimat_s_2} \\
 | \rho_{i\,j\,k} - \rho_{i\,j\,k+1} | & \leq & c_s \quad \forall \; i \leq n_x,   \; j \leq n_y,   \; k \leq n_z-1, \label{eqn:prob_bimat_s_3} \\ 
\rho_e & \in & [0,1] \quad \forall  \; 1 \leq e \leq N. \label{eqn:prob_bimat_mat} 
\end{eqnarray}
Here we effectively have a single variable $\rho_e$ for each of our $N$ design cells, see \secref{sec:parametrization}. Moreover, volume, grayness and slope constraints act on $\rho_e$ only.   

In the full three-material optimization model we have $3\,N$ design variables $\rho_1^A$, $\rho_1^B$, $\rho_1^C$, $\rho_2^A$,  $\ldots$, $\rho_N^C$, but it is sufficient to double constraints \eqnref{eqn:prob_bimat_v}--\eqnref{eqn:prob_bimat_mat} for a second variable only, because  the third one is implicitly controlled. With the additional local  condition \eqnref{eqn:rho_j_L1_norm}, we obtain the full three-material problem formulation
\begin{eqnarray}
\max_\brho && J(\brho, \bh) \Big \vert_{t=T}  \label{eqn:prob_trimat_J} \\
\text{s.t. } \bM(\brho, \bh)\,\dt{\bh} - \bK(\brho, \bh)\,\bh  & = & \bzero, \label{eqn:prob_trimat_state}\\
v^A(\brho) & = & v_A^*, \label{eqn:prob_trimat_v_A} \\
v^C(\brho) & = & v_C^*, \label{eqn:prob_trimat_v_C} \\
g^A(\brho) & \leq & c_g^A,  \label{eqn:prob_trimat_g_A} \\
g^C(\brho) & \leq & c_g^C,  \label{eqn:prob_trimat_g_C} \\
 | \rho^A_{i\,j\,k} - \rho^A_{i+1\,j\,k} | & \leq & c_s \quad \forall \; i \leq n_x-1, \; j \leq n_y,   \; k \leq n_z,  \label{eqn:prob_trimat_s_1} \\
 | \rho^A_{i\,j\,k} - \rho^A_{i\,j+1\,k} | & \leq & c_s \quad \forall \; i \leq n_x,   \; j \leq n_y-1, \; k \leq n_z, \\
 | \rho^A_{i\,j\,k} - \rho^A_{i\,j\,k+1} | & \leq & c_s \quad \forall \; i \leq n_x,   \; j \leq n_y,   \; k \leq n_z-1, \\ 
 | \rho^C_{i\,j\,k} - \rho^C_{i+1\,j\,k} | & \leq & c_s \quad \forall \; i \leq n_x-1, \; j \leq n_y,   \; k \leq n_z, \\
 | \rho^C_{i\,j\,k} - \rho^C_{i\,j+1\,k} | & \leq & c_s \quad \forall \; i \leq n_x,   \; j \leq n_y-1, \; k \leq n_z, \\
 | \rho^C_{i\,j\,k} - \rho^C_{i\,j\,k+1} | & \leq & c_s \quad \forall \; i \leq n_x,   \; j \leq n_y,   \; k \leq n_z-1, \label{eqn:prob_trimat_s_6} \\ 
\rho_e^A + \rho_e^B + \rho_e^C & = & 1 \quad \forall  \; 1 \leq e \leq N,  \label{eqn:prob_trimat_L1_norm} \\ 
\rho_e^A & \in & [0,1] \quad \forall  \; 1 \leq e \leq N,  \label{eqn:prob_trimat_mat_A} \\ 
\rho_e^B & \in & [0,1] \quad \forall  \; 1 \leq e \leq N,  \label{eqn:prob_trimat_mat_B} \\
\rho_e^C & \in & [0,1] \quad \forall  \; 1 \leq e \leq N.  \label{eqn:prob_trimat_mat_C}
\end{eqnarray}
It is noted that the third material variable could be expressed as $\rho_e^B = 1 - \rho_e^A - \rho_e^C$, but this would only reduce the number of variables and not the number of constraints. In topology optimization, the number of constraints  is more critical in terms of numerical performance.  

\section{Results of the bi-material topology optimization}\label{sec:results_bi}

For simplicity, we began with a bi-material optimization problem, whose formulation and presentation of the results are much simpler than for the three-material problem. In this section, the spatial distributions of materials $A$ and $C$ were optimized, whereas no buffer material ($B$) was used in the designs. The corresponding two-layer reference design consisted here of a layer of material $A$ (25\% volume fraction) directly on top of a layer of material $C$ (75\% volume fraction). As for the reference three-layer design described in \secref{sec:reference}, the liquid discharge lasted for 24\,s and this was followed by an equilibration phase of 120\,s. With this setting, the reference two-layer design absorbed $64.5$\,cm$^3$, which is much more than the $41$\,cm$^3$ of the reference three-layer design, where $A$ has a volume fraction of only 5\% and $B$ 10\%. Clearly, increasing the volume  of the most permeable material ($A$) results in a much larger volume of absorbed liquid.

\subsection{Unregularized Problem}\label{sec:bimat_unreg}

\begin{figure}[!ht]
 \centering
 \subfloat[continuous optimized design ]
 {\includegraphics[width=.45\textwidth]{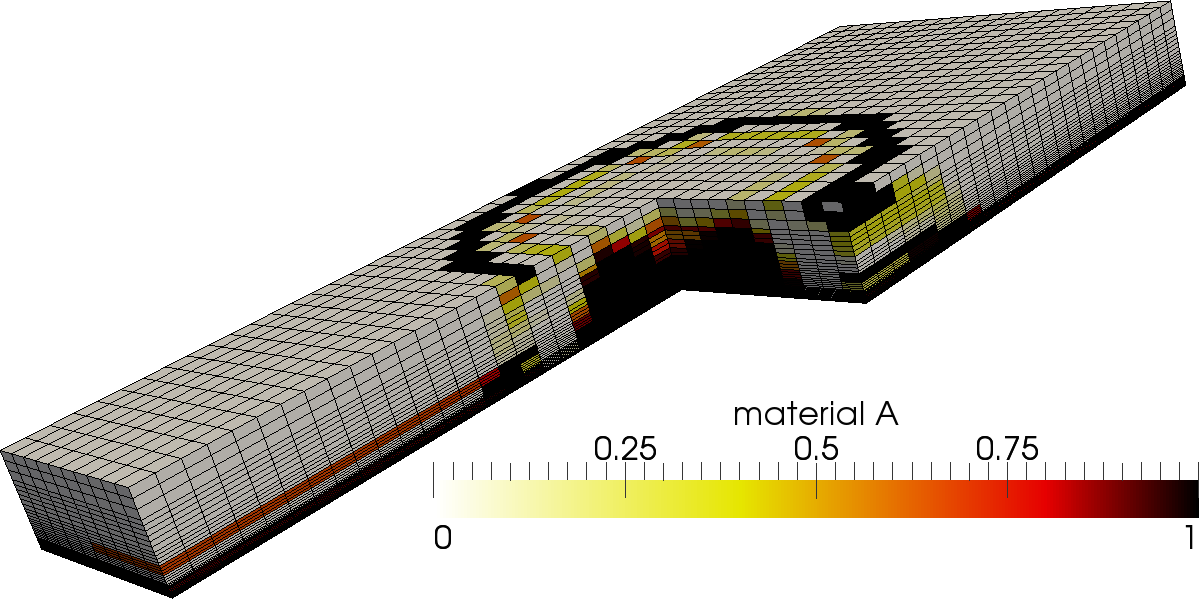} \label{fig:bimat_mat_a}}
 \subfloat[binary design] 
 {\includegraphics[width=.45\textwidth]{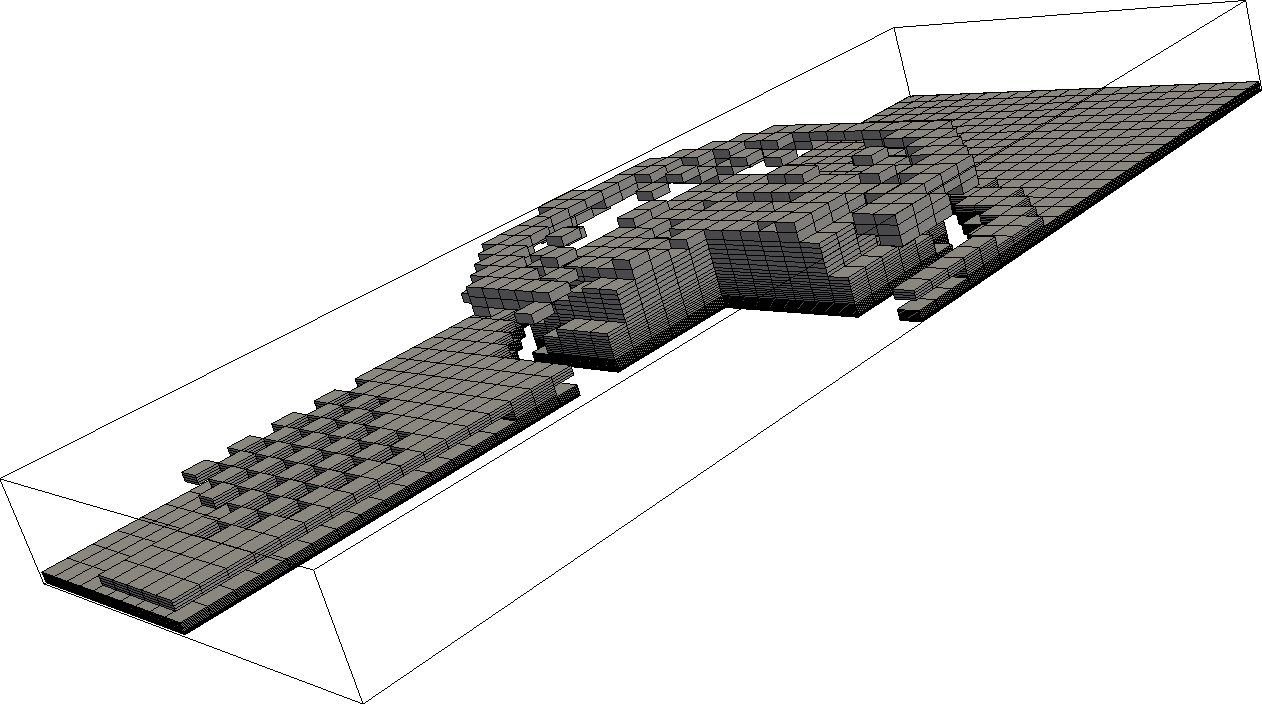} \label{fig:bimat_mat_a_th}} 
  \caption{\label{fig:bimat_simple}Results of the bi-material optimization, in which the distributions of materials $A$ and $C$ are varied, whereas material $B$ is not used. Here 25\% of the volume is filled with material $A$ and 75\% with $C$. No slope and grayness constraints are applied. All visualizations are upside down, with the liquid discharge from below, and one quarter of the domain is omitted to better show the inner structure of the diaper. (a)  Material $A$ is shown as black, whereas $C$ is shown as gray. Mixed $A$-$C$ material (shown in color) appears in the optimal design.  (b) Material is rounded to discard mixed material. To enhance the visualization, material  $A$ is shown in gray (instead of black), whereas material $C$ is not shown. The volume fractions are preserved by rounding.}
\end{figure}

The first problem was posed without grayness and slope constraints, hence the problem consisted of equations \eqnref{eqn:prob_bimat_J}, \eqnref{eqn:prob_bimat_state}, \eqnref{eqn:prob_bimat_v} and \eqnref{eqn:prob_bimat_mat}. The optimized design is shown in \figref{fig:bimat_mat_a}, where the discharge is from below to aid visualization of the structure.  In this design, material $A$ is removed from the areas far away from the discharge inlet and is placed below it, forming a thick layer of complex form.  Because of the high permeability of material $A$ this design allows evacuating the fluid faster from the discharge inlet, thereby increasing the flux of liquid into the diaper. This leads to a substantially better performance. Compared to the initial design, the total absorbed volume (objective function) is increased by $45\%$ ($64.5 \to 93.4$cm$^3$). Note that with a volume fraction of $25\,\%$ for material $A$, not all of it is necessary and as a result part of it remains on the top (bottom in the visualization) or is moved to other regions which remain dry during the whole simulation time. The obtained result shows some mixed material, especially close to the discharge area. For this reason, we performed a material-fraction preserving rounding, which led to the material distribution in \figref{fig:bimat_mat_a_th}. Interestingly, this rounding has almost no impact on the objective value ($93.4 \to 93.2$\,cm$^3$).

\subsection{Regularized Problem}

\begin{figure}[!ht]
 \centering
 \subfloat[slope constraints]
 {\includegraphics[width=.45\textwidth]{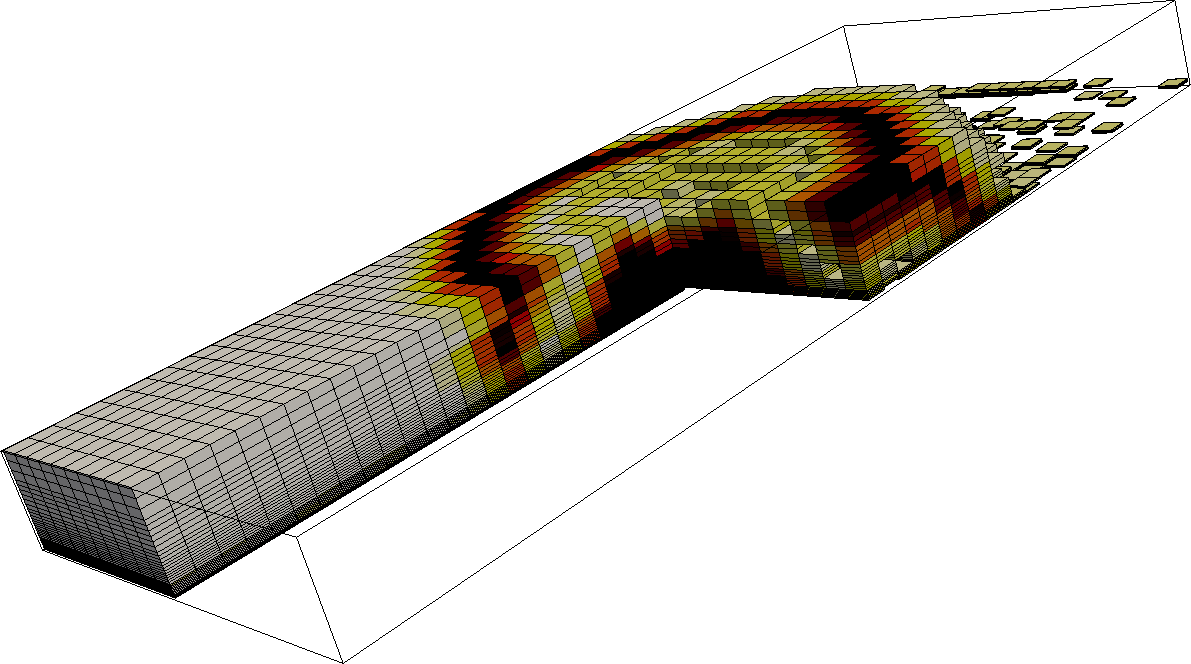} \label{fig:bimat_slope}}
 \subfloat[additional grayness constraints] 
 {\includegraphics[width=.45\textwidth]{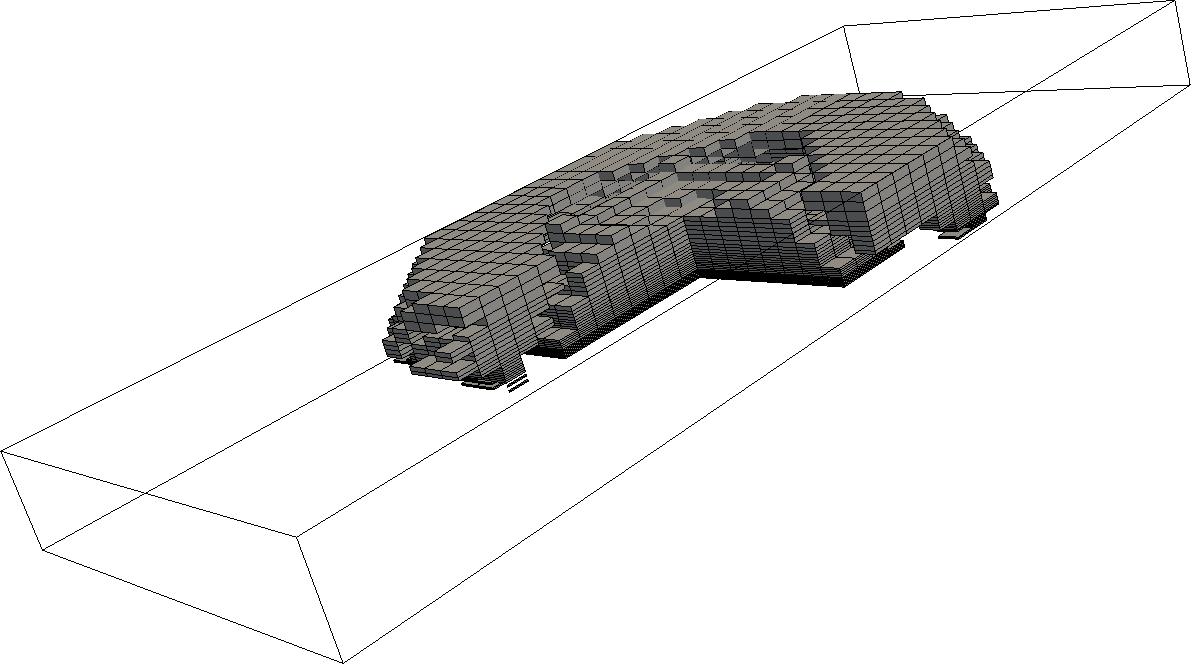} \label{fig:bimat_slope_gray}} 
  \caption{\label{fig:bimat_slope_grayness}(a): The bi-material problem shown in \figref{fig:bimat_simple} is repeated with additional slope constraints. (b): Design after additional grayness constraints and rounding to distinguish materials $A$ (shown as gray ) and $C$ (not shown).}
\end{figure}

As a second problem, we added the slope constraints \eqnref{eqn:prob_bimat_s_1}--\eqnref{eqn:prob_bimat_s_3} to the bi-material problem. The slope parameter was set to $c_s=0.3$, hence approximately two intermediate elements were necessary to change from pure material $A$ to $C$ and vice versa. The resulting design is shown in \figref{fig:bimat_slope} and is similar to that of \figref{fig:bimat_mat_a}. However,  because of the imposed slower transition between materials, there is more mixed material than enforced by the slope constraints and there is significantly less material $A$ left over on the top layer (bottom in the visualization) far away from the discharge inlet. The restriction of the design space by slope constraints leads to an objective function value of $91.1$\,cm$^3$, which is still an improvement of $41\%$ compared to the layered reference design.

Next, we calculated the grayness induced by the slope constraints for the design shown in \figref{fig:bimat_slope}, as described in \secref{sec:functions}, and added the value as bound for the additional grayness constraint \eqnref{eqn:prob_bimat_g}. We solved this second problem starting from the solution obtained from the slope constraint reference problem. The resulting binary design is shown in \figref{fig:bimat_slope_gray}. The additional grayness function has almost no impact on the obtained value of the objective function. For the binary design the volume of absorbed liquid even increases slightly to $92.5$\,cm$^3$.
This is an indication that intermediate material, here enforced by the slope constraints, is obviously not necessarily advantageous for such bi-material problems.  

\subsection{Effect of the liquid discharge and equilibration times}

\begin{figure}[!ht]
 \centering
 \subfloat[short discharge]
 {\includegraphics[width=.45\textwidth]{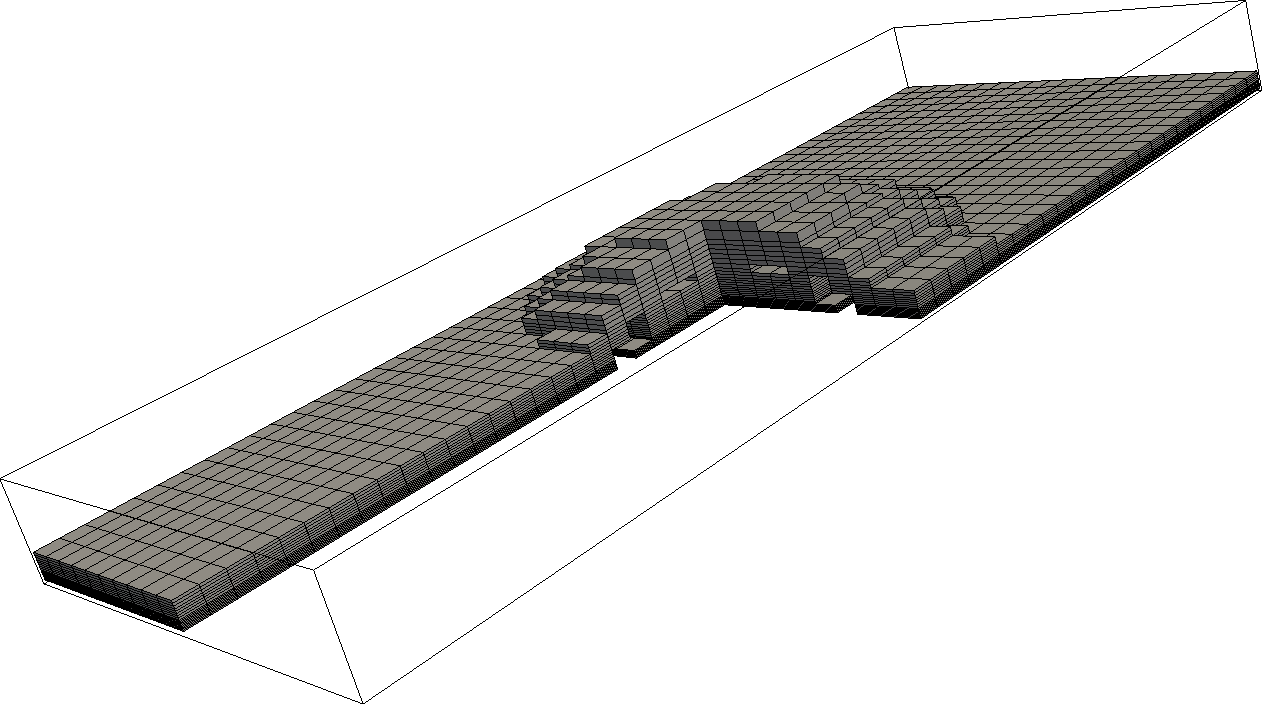} \label{fig:bimat_short}}
 \subfloat[long discharge] 
 {\includegraphics[width=.45\textwidth]{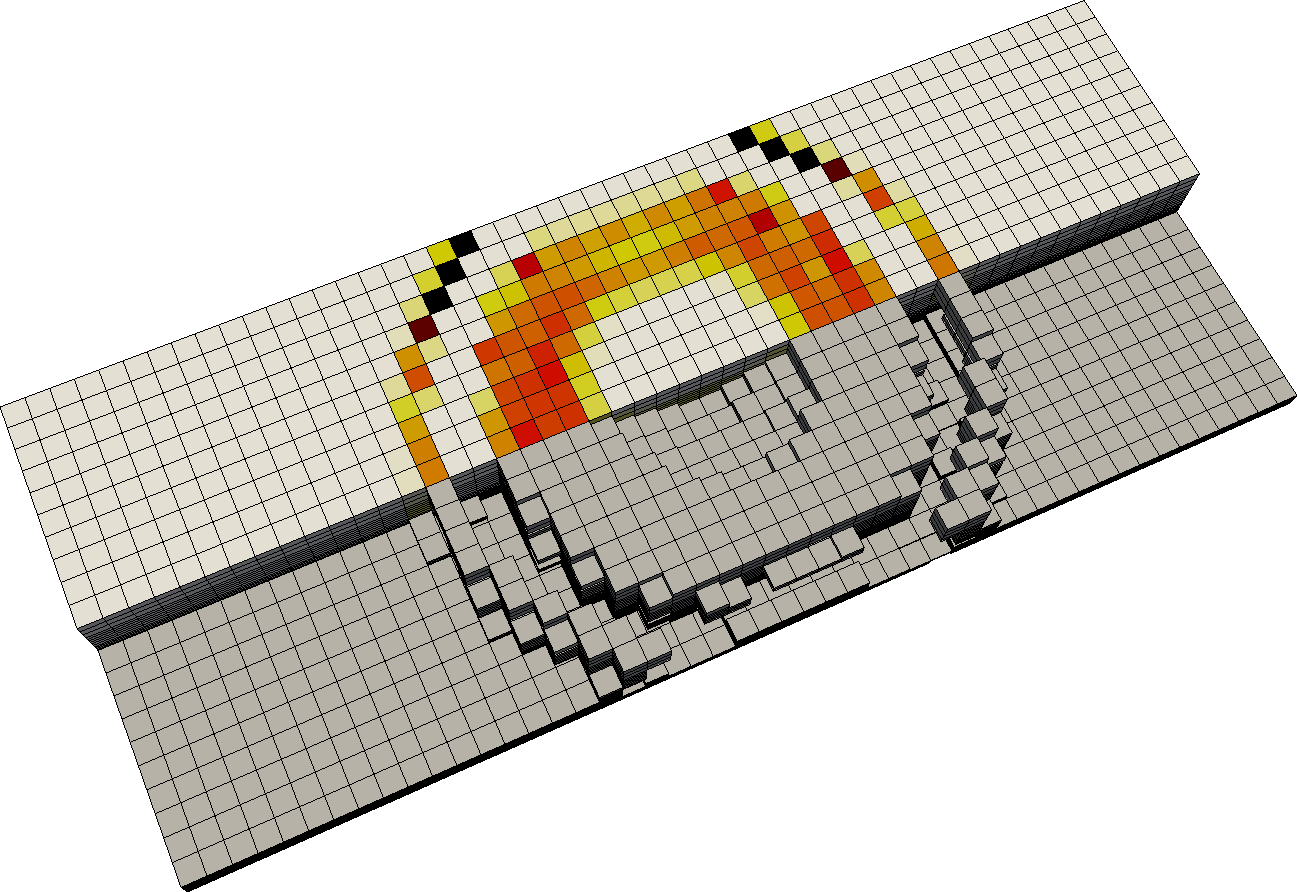} \label{fig:bimat_long}} 
  \caption{\label{fig:bimat_timing}Bi-material optimization with a variation of the discharge time. (a) With a short discharge of 3\,s, the obtained design is almost perfectly free of intermediate material. Material $A$ is therefore shown in gray. (b) Long discharge of 48\,s followed by 240\,s of equilibration time compared to the standard case. The upper part of the structure uses the same visualization scheme as in  \figref{fig:bimat_mat_a} (where $A$ is black, $C$ gray and color corresponds to mixed material, whereas the lower part is visualized as in \figref{fig:bimat_mat_a} (where only material $A$ is shown).}
\end{figure}

Next, we studied the effect of changing the duration of the discharge and equilibration phases. First, we reduced the discharge time to 3\,s and the  equilibration time to 27\,s. For this problem, regularization was not necessary. We furthermore skipped the resource constraint such that the problem reduces to \eqnref{eqn:prob_bimat_J}, \eqnref{eqn:prob_bimat_state} and \eqnref{eqn:prob_bimat_mat}. The distribution of material $A$ of the obtained design is shown in \figref{fig:bimat_short}. This design has practically no mixed material and is therefore shown in gray to aid visualization. The objective function only increases from $23.9$\,cm$^3$ in the two-layer initial design with material $A$ filling $25\%$ of the domain, to an objective value of $26.7$\,cm$^3$ with material $A$ filling $30\%$ of the domain. As in the previous cases, the design shows a bubble of material $A$ below the discharge inlet, but appears simpler in topology. The initial layer of material $A$  at the top remains nearly untouched as it is remains dry within the short simulation time.         
 
Finally, we considered a long discharge of 48\,s followed by an equilibration phase of 240\,s. We again show the design obtained  without resource constraint and without slope constraints in \figref{fig:bimat_long}. There is a centered bubble of material $A$, with a small dent filled with material $C$. The bubble is surrounded by a ring of material $C$, which is itself surrounded by a thin ring of material $A$. Beyond this described structure, the design is in its initial state in regions which remain dry. The objective function increases from 118 to 176\,cm$^3$. Solving the same problem with material constraint and with or without slope constraints, yields a similar design. The presented one visualizes the design principle best.  

\section{Three-Material Optimization}\label{sec:results_three}

\begin{figure}[!ht]
 \centering
 \subfloat[unregularized]
 {\includegraphics[width=.45\textwidth]{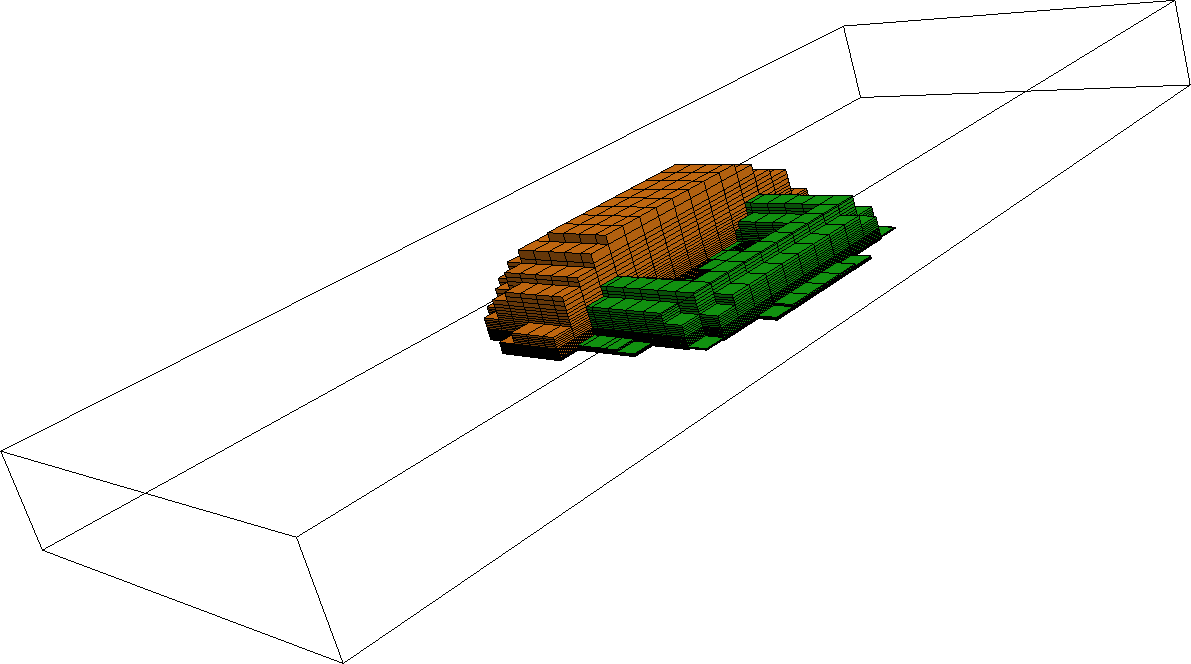} \label{fig:opt_three}}
 \subfloat[slope constraints] 
 {\includegraphics[width=.45\textwidth]{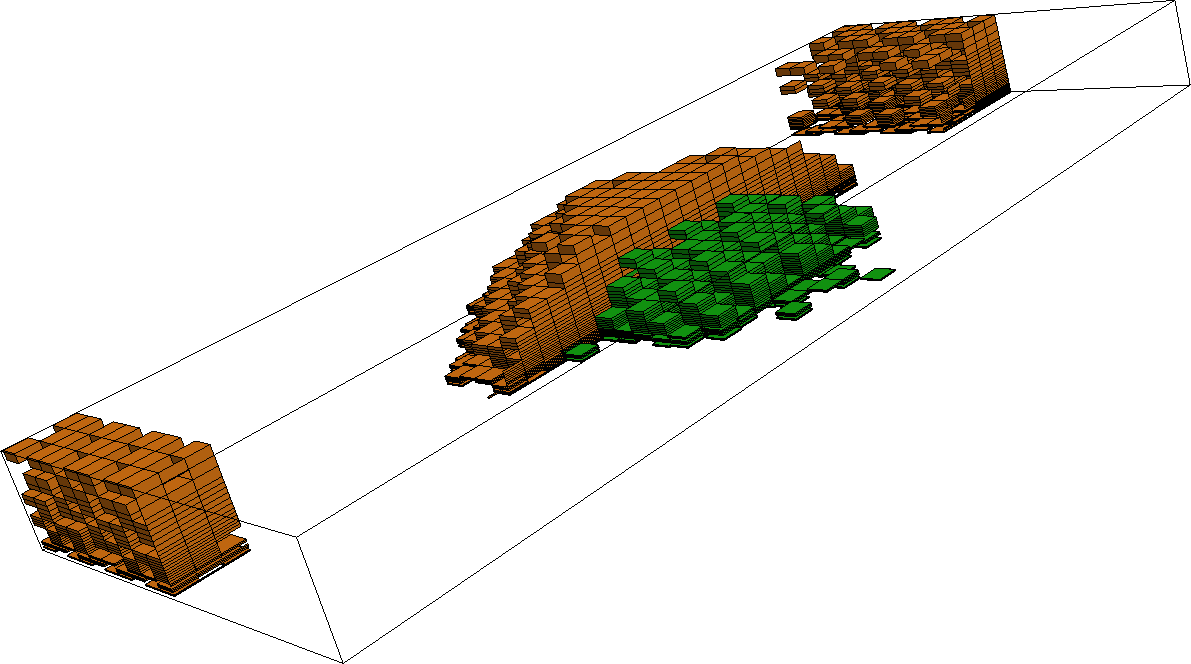} \label{fig:opt_three_slope}} 
  \caption{\label{fig:three_mat} Designs obtained with three-material optimization. Material $A$ is shown in green, material $B$ in orange and the remaining material ($C$) is not shown. (a) Without slope constraints the obtained design is clear with distinct material. (b) Enforcing material mixes by slope constraints leads to an indistinct design with poor performance.}
\end{figure}

We here focus on the optimization of the reference three-layer design described in \secref{sec:reference}, where material $A$ is restricted to $5\%$ \eqnref{eqn:prob_trimat_v_A}, material $C$ to $85\%$ \eqnref{eqn:prob_trimat_v_C} and the same liquid discharge along with equilibration times as for the standard bi-material problem is used. Here the spatial distribution of the buffer material occupies 10\% of the volume and its distribution is allowed to change as well. The problem without regularization and grayness constraints is given by the equations \eqnref{eqn:prob_trimat_J} -- \eqnref{eqn:prob_trimat_v_C} and \eqnref{eqn:prob_trimat_L1_norm}--\eqnref{eqn:prob_trimat_mat_C}. The obtained design features almost no mixed materials and is shown in \figref{fig:opt_three}. The objective function increases here from 41 to 86\,cm$^3$, which is achieved with structures reminiscent of those found in the bi-material optimization, but with important differences. Firstly, because material A (shown as green) is only 5\% of the total volume, it cannot be employed to form a large bubble below the discharge inlet. Instead, the bubble of material $A$ is much smaller and is surrounded by a layer along the border of the inlet, which is thinner in the horizontal direction, while penetrating deeper. This allows to distribute liquid efficiently along the horizontal direction. Secondly, the central region of the bubble, which was formed of material $A$ in the bi-material designs, is now made of material $B$ (shown as orange). Although $B$ is less permeable than $A$, it is still much more permeable than $C$. 

Adding regularization by slope constraints with $c_s=0.3$, we have with the exception of grayness constraints the full problem formulation \eqnref{eqn:prob_trimat_J} -- \eqnref{eqn:prob_trimat_v_C} and \eqnref{eqn:prob_trimat_s_1}--\eqnref{eqn:prob_trimat_mat_C}. The enforced material mix is clearly disadvantageous, the design becomes much less distinct as depicted in \figref{fig:opt_three_slope} and the obtained objective function value reduces to 74\,cm$^3$. 
 
\subsection{Variants}
\label{sec:variants}

We carried out numerical simulations with extended problem formulations, which we summarize in what follows. A real diaper must not only absorb as much liquid as possible during the discharge, but should also keep the top layer, which is in contact with the skin, as dry as possible. The function for the moisture content \eqnref{eqn:f_total} can easily be defined for a subarea $\Omega_q$ of the complete computational domain $\Omega$, together with an own adjoint problem to be solved. We defined a top layer of 0.3\,mm and limited the amount of fluid within the top layer by a constraint function in the sense of a wetness constraint. The result is that the top layer is filled with material $A$ and partly with $B$ (far away from the discharge inlet) because they have a small  moisture capacity. Except for this difference, the optimized design remain similar. 

A further point is that several liquid discharges may occur before the diaper is discarded. We performed optimization runs with several discharge and equilibration phases and again no significant impact on the design was observed. Finally, we considered several  discharge locations associated with a discharge probability. This requires the separate solution of the state and adjoint problems for each location. For the objective function, the discharges were combined as a weighted sum (with the probability above) and a separate wetness constraint was associated to each discharge case. This complicates the problem formulation further. Material $A$ and $B$ need to be distributed to cover all load locations. The resulting design applies the general design principle according to their weighting factor for the individual load cases and is less clear. As a result, the gains in performance are much more limited, only of about 6\%, in comparison to the reference three-layer design.

\section{Discussion and Conclusion}
\label{sec:conclusion}

A model for gradient based topology optimization of different materials within an unsaturated flow problem was presented. Our benchmark problem was a  highly simplified computational model of liquid propagation in a diaper, where the spatio-temporal evolution of the moisture content is obtained from the solution of Richards' equation. A versatile finite-volume solver was implemented in OpenFOAM for this purpose and was further extended to solve the adjoint equation resulting from the topology optimization. It was shown that gravitational force plays a negligible role in the imbibition process, which is mainly dominated by capillary force.

Our topology optimization yielded designs showing a substantially enhanced capability to absorb liquid with respect to a simple layered design. The optimized designs are similar in all investigated bi-material and three-material problems. Their main common feature is the clustering of highly permeable material close to the discharge inlet in order to allow a fast evacuation of the liquid from the surface. This results in a higher flux of liquid through the inlet and hence a larger total volume is absorbed at the end of the simulation. 

All the bi-material optimized designs differ only in subtle details and present a similar performance. This is because of the large volume fraction of the most permeable material $A$ considered (25\%). In fact, in most cases there is so much volume of $A$ that parts of it are left on the top of the diaper, in the regions which never hold any moisture. In all cases, a large bubble of material $A$ is formed and reaches from the discharge inlet at the top, to the bottom of the diaper. Because of the large permeability of $A$, these designs enable a rapid distribution of the liquid in the horizontal direction, where the liquid is then stored in the  superabsorber (material $C$).    

The three-material problem considered here is closer to a real diaper. In this case, only 5\% of the total volume is composed of material $A$, 10\% of the buffer material $B$ and the remaining 85\% of the superabsorber $C$. The large bubble of high permeability below the inlet remains similar in shape and size, but because of the scarcity of material $A$, the bubble is composed not only of material $A$, but also of the buffer material $B$. As expected, $A$ forms a layer immediately below the inlet, which is slightly thicker than the one of the layered reference design. However, in a narrow region at the perimeter of the inlet, this layer of material $A$ protrudes down to the bottom of the diaper. This enhances the distribution of the liquid far away from the inlet in the horizontal direction. Material $B$ covers $A$ completely and reaches down to the bottom of the diaper. Therefore the role of material $B$ in the optimized design is similar to its role in the reference three-layer design, namely to act as a transition region between $A$ and $C$. 

Very small volumes of mixed material appeared in the optimized designs. This can be explained by its poor performance. In fact, regions of mixed material occurred only in the transition between pure materials and merely because the volume fraction of each material must be preserved in the optimization. Hence, we expect that with finer discretization, the volume of mixed materials would be even smaller. At the end of the optimization procedure, mixed material was eliminated with a volume-preserving rounding, which resulted in a negligible loss of performance. Regularization of the problem with slope constraints led to worse results.

Finally, we note that some of the fine features of the optimized designs are not obvious. A detailed understanding of the underlying mechanisms and their relationship to the spatial and temporal settings and material properties, is an interesting question in itself, which is beyond the scope of this publication. In addition, the model may be adopted for arbitrary applications of unsaturated flow problems in heterogeneous porous materials. The objective here was to maximize the amount of absorbed fluid, but with the presented model, more complex problems with upper and lower bounds for the amount of absorbed fluids in different regions, or more complex discharge cases, can be easily formulated.

\section*{Acknowledgements}

The authors gratefully acknowledge the support of the Cluster of Excellence 'Engineering of Advanced Materials' at the University of Erlangen-Nuremberg, which is funded by the German Research Foundation (DFG) within the framework of its 'Excellence Initiative'.

\bibliography{papers_opt,sim}

\end{document}